\documentclass[twocolumn,floatfix,superscriptaddress]{revtex4-1}
\pdfoutput=1
\usepackage{graphicx}
\usepackage{times}
\usepackage{amsmath,amssymb,bm}
\usepackage{dcolumn}
\usepackage{mathtools}
\usepackage{color}
\usepackage{mathrsfs}
\usepackage{txfonts}
\newcommand{\ket}{\rangle}
\newcommand{\bra}{\langle}

\usepackage{dsfont}
\newcommand{\pkpr}{P_3}

\usepackage{subfigure}

\begin{document}

\title{A strongly correlated metal built from Sachdev-Ye-Kitaev models}
\author{Xue-Yang Song}
\affiliation{International Center for Quantum Materials, School of Physics, Peking University, Beijing, 100871, China}\affiliation{Kavli Institute of Theoretical Physics, University of California, Santa Barbara, CA 93106, USA}
\author{Chao-Ming Jian}
\affiliation{Kavli Institute of Theoretical Physics, University of California, Santa Barbara,  CA 93106, USA}
\affiliation{Station Q, Microsoft Research, Santa Barbara, California 93106-6105, USA}
\author{Leon Balents}
\affiliation{Kavli Institute of Theoretical Physics, University of California, Santa Barbara, CA 93106, USA}

\begin{abstract}
  {\bf Strongly correlated metals comprise an enduring puzzle at the heart of condensed matter physics.  Commonly a highly renormalized heavy Fermi liquid occurs below a small coherence scale, while at higher temperatures a broad incoherent regime pertains in which quasi-particle description fails.  Despite the ubiquity of this phenomenology, strong correlations and quantum fluctuations make it challenging to study. The Sachdev-Ye-Kitaev(SYK) model describes a $0+1$D quantum cluster with random all-to-all \emph{four}-fermion interactions among $N$ Fermion modes which becomes exactly solvable as $N\rightarrow \infty$, exhibiting a zero-dimensional non-Fermi liquid with emergent conformal symmetry and complete absence of quasi-particles.  Here we study a lattice of complex-fermion SYK dots with random inter-site \emph{quadratic} hopping.  Combining the imaginary time path integral with \emph{real} time path integral formulation, we obtain a heavy Fermi liquid to incoherent metal crossover in full detail, including thermodynamics, low temperature Landau quasiparticle interactions, and both electrical and thermal conductivity at all scales.  We find linear in temperature resistivity in the incoherent regime, and a Lorentz ratio $L\equiv \frac{\kappa\rho}{T}$ varies between two universal values as a function of temperature. Our work exemplifies an analytically controlled study of a strongly correlated metal.}
\end{abstract}

\date{\today}

\maketitle

Prominent systems like the high-T$_c$ cuprates and heavy fermions display intriguing features going beyond the quasiparticle description\cite{emery_1995,emery_19952,varma_1996,mathur_1998,zaanen_2004,hartnoll_20152,hartnoll_2015,bruin_2013,stewart_2001}. 
The exactly soluble SYK models provide a powerful framework to study such physics.  The most-studied SYK$_4$ model, a $0+1$D quantum cluster of $N$ Majorana fermion modes with random all-to-all four-fermion interactions\cite{sachdev_1993, kitaev_kitp,sachdev_2015,maldacena_2016,polchinski_2016,fu_20162,you_2017,gu_2017,sannomiya_2017} has been generalized to SYK$_q$ models with $q$-fermion interactions.
Subsequent works\cite{gu_2016,davidson_2016} extended the SYK model to higher spatial dimensions by coupling a lattice of SYK$_{4}$ quantum clusters by additional four-fermion ``pair hopping'' interactions.   They obtained electrical and thermal conductivities completely governed by diffusive modes and nearly temperature-{\em independent} behavior owing to the identical scaling of the inter-dot and intra-dot couplings.  

Here, we take one step closer to realism by considering a lattice of complex-fermion SYK clusters with SYK$_4$ intra-cluster interaction of strength $U_0$ and random inter-cluster  ``SYK$_2$" two-fermion hopping of strength $t_0$\cite{kolovsky_2017,jian_2017,bi_2017,banerjee_2016,jian2_2017}.  Unlike the previous higher dimensional SYK models where local quantum criticality governs the entire low temperature physics, here as we vary the temperature, two distinctive metallic behaviors appear,  resembling the previously mentioned heavy fermion systems. We assume $t_0 \ll U_0$, which implies strong interactions, and focus on the correlated regime $T \ll U_0$.   We show the system has a coherence temperature scale $E_c \equiv t_0^2/U_0$\cite{jacquod_1997,kolovsky_2017,ho_1998} between a  heavy Fermi liquid  and an incoherent metal.   For $T < E_c$, the SYK$_2$ induces a Fermi liquid, which is however highly renormalized by the strong interactions.  For $T> E_c$, the system enters the incoherent metal regime and the resistivity $\rho$ depends linearly on temperature.   These results are strikingly similar to those of Parcollet and Georges\cite{parcollet_1999}, who studied a variant SYK model obtained in a double limit of infinite dimension and large $N$.   Our model is simpler, and does not require infinite dimensions.   We also obtain further results on the thermal conductivity $\kappa$, entropy density and Lorentz ratio\cite{franz_1853,sommerfeld_1927} in this crossover.  This  work bridges traditional Fermi liquid theory and the hydrodynamical description of an incoherent metallic system.

\emph{SYK model and Imaginary-time formulation - }  We consider a $d$-dimensional array of quantum dots, each with  $N$ species of fermions labeled by $i,j,k\cdots$, 
\begin{eqnarray}
\label{hamiltonian}
&&\mathcal H=\sum_x \sum_{i<j,k<l} U_{ijkl,x}c_{ix}^\dagger c_{jx}^\dagger c^{\vphantom\dagger}_{kx}c^{\vphantom\dagger}_{lx}+\sum_{\bra xx'\ket} \sum_{i,j}t_{ij,xx'} c_{i,x}^\dagger c^{\vphantom\dagger}_{j,x'}
\end{eqnarray}
where $U_{ijkl,x}=U_{klij,x}^*$ and $t_{ij,xx'}=t_{ji,x'x}^*$ are random zero mean complex variables drawn from Gaussian distribution whose variances $\overline{|U_{ijkl,x}|^2} = 2 U_0^2/N^3$ and $\overline {|t_{ij,x,x'}|^2} = t_0^2/N$.  

In the imaginary time formalism, one studies the partition function $Z={\rm Tr}\, e^{-\beta (\mathcal{H}-\mu \mathcal{N})}$,  with $\mathcal{N} = \sum_{i,x} c_{i,x}^\dagger c_{i,x}^{\vphantom\dagger}$, written as a path integral over Grassman fields  $c_{ix\tau},\bar c_{ix\tau}$.  Owing to the self-averaging established for the SYK model at large $N$, it is sufficient to study $\bar{Z}= \int [d\bar{c}][dc] e^{-S_c}$, with (repeated species indices are summed over)
\begin{widetext}
  \begin{eqnarray}
    \label{eq:1}
    S_c & = & \sum_{x}\int_0^\beta \! d\tau \, \bar c_{ix\tau}(\partial_\tau - \mu) c_{ix\tau} - \int_0^\beta \! d\tau_1 d\tau_2 \,\Big[ \sum_x\frac{U_0^2}{4N^3} \,\bar{c}_{ix\tau_1} \bar{c}_{jx\tau_1}  c_{kx\tau_1} c_{lx\tau_1} \bar{c}_{lx\tau_2} \bar{c}_{kx\tau_2} c_{jx\tau_2} c_{ix\tau_2}  +\sum_{\bra xx'\ket} \frac{t_0^2}{N} \, \bar{c}_{ix\tau_1} c_{jx'\tau_1} \bar{c}_{jx'\tau_2} c_{ix\tau_2} \Big].
  \end{eqnarray}
\end{widetext}
The basic features can be determined by a simple power-counting.  Considering for simplicity $\mu=0$, starting from $t_0=0$, the $U_0^2$ term is invariant under $\tau \rightarrow b \tau$ and $c \rightarrow b^{-1/4} c$, $\bar{c}\rightarrow b^{-1/4} \bar{c}$, fixing the scaling dimension $\Delta=1/4$ of the fermion fields.  Under this scaling $\bar{c}\partial_\tau c$ term is irrelevant. Yet upon addition of two-fermion coupling, under rescaling, $t_0^2 \rightarrow b t_0^2$, so two-fermion coupling is a {\em relevant} perturbation.  By standard reasoning, this implies a cross-over from the SYK$_4$-like model to another regime at the energy scale where the hopping perturbation becomes dominant, which is $E_c = t_0^2/U_0$.  We expect the renormalization flow is to the SYK$_2$ regime.  Indeed keeping the SYK$_2$ term invariant fixes $\Delta = 1/2$, and $U_0^2 \rightarrow b^{-1} U_0^2$ is irrelevant.  Since the SYK$_2$ Hamiltonian (i.e.,$U_0=0$) is quadratic, the disordered free fermion model supports quasi-particles and defines a Fermi liquid limit.  For $t_0 \ll U_0$, $E_c$ defines a crossover scale between SYK$_4$-like non-Fermi liquid and the low temperature Fermi liquid.

\begin{figure}[htbp] 
 \begin{center}
  \includegraphics[width=0.40\textwidth]{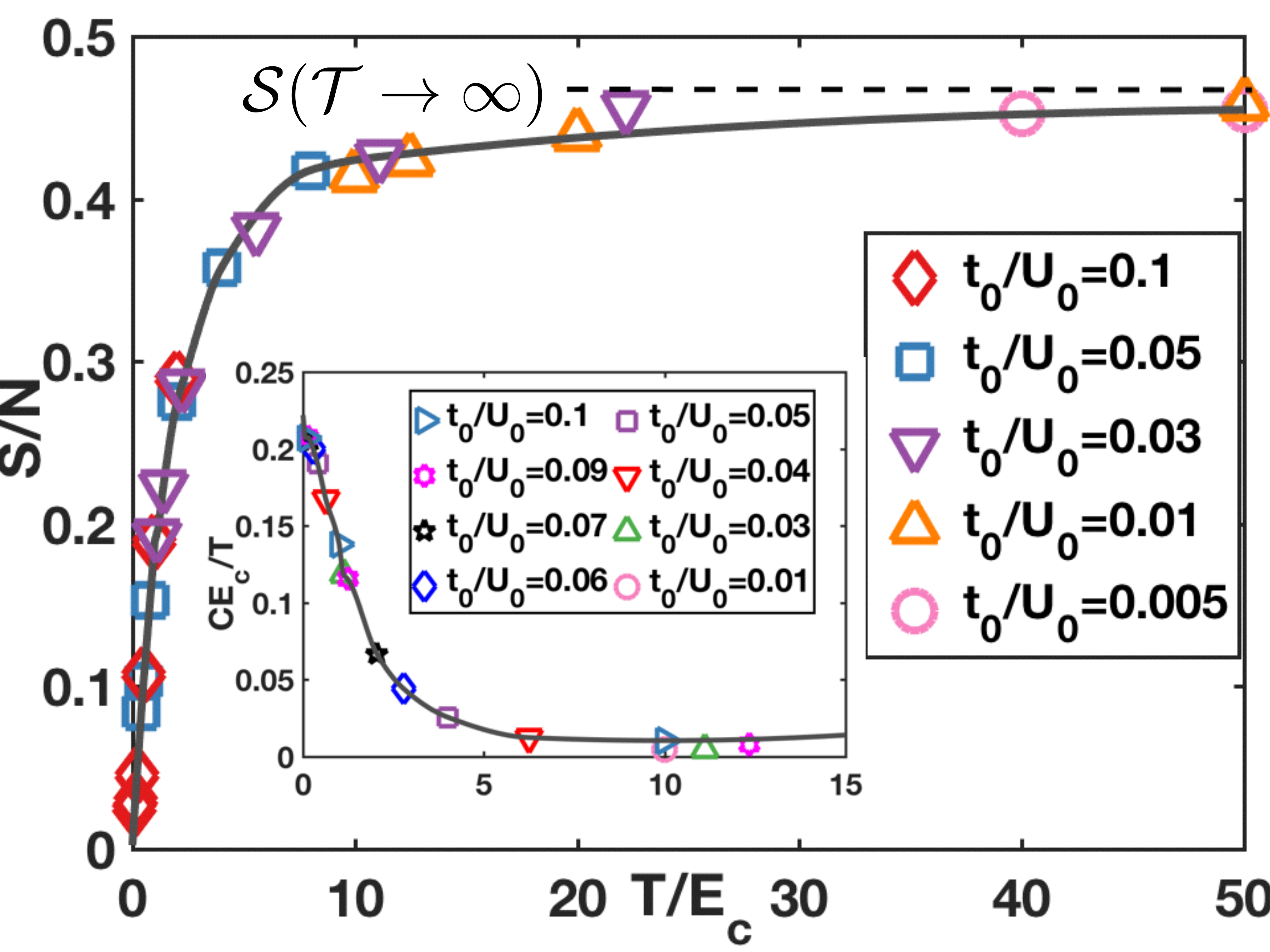}
 \caption{The entropy and specific heat(inset) collapse to universal functions of $\frac{T}{E_c}$, given $t_0,T\ll U_0$($z=2$).  $C\rightarrow \mathcal S'(0)T/E_c$ as $T/E_c\rightarrow 0$. Solid curves are guides to the eyes.
  }    \label{thermo}
 \end{center}
 \end{figure}

At the level of thermodynamics, this crossover can be rigorously established using imaginary time formalism.  Introducing a composite field $G_x(\tau,\tau')=\frac{-1}{N}\sum_i c_{ix\tau}\bar c_{ix\tau'}$ and a Lagrange multiplier $\Sigma_x(\tau,\tau')$ enforcing the previous identity, one obtains $\bar{Z} = \int [d G][d\Sigma] e^{-N S}$, with the action
\begin{widetext}
  \begin{eqnarray}
\label{euc_action_fermion}
 &&S= - \sum_x \ln \det \left[(\partial_\tau-\mu) \delta(\tau_1-\tau_2)+\Sigma_x(\tau_1,\tau_2)\right]
 + \int_0^\beta \! d\tau_1 d\tau_2 \Big( - \sum_x \left[ \frac{U_0^2}{4}G_x(\tau_1,\tau_2)^2G_x(\tau_2,\tau_1)^2 +\Sigma_x(\tau_1,\tau_2)G_x(\tau_2,\tau_1)\right]
\nonumber \\
&& +t_0^2\sum_{\langle xx'\rangle} G_{x'}(\tau_1,\tau_2)G_x(\tau_2,\tau_1) \Big).
\end{eqnarray}
\end{widetext}
The large $N$ limit is controlled by the saddle point conditions $\delta S/\delta G = \delta S/\delta\Sigma = 0$, satisfied by $G_x(\tau,\tau') = G(\tau-\tau')$, $\Sigma_x(\tau,\tau') = \Sigma_4(\tau-\tau')+ zt_0^2 G(\tau-\tau')$ ($z$ is the coordination number of the lattice of SYK dots), which obey 
\begin{eqnarray}
\label{SDm}
G(i\omega_n)^{-1} & = & i\omega_n+\mu-\Sigma_{4}(i\omega_n)-zt_0^2G(i\omega_n),\nonumber\\
\Sigma_4 (\tau) & = & -U_0^2G(\tau)^2G(-\tau),
\end{eqnarray}
where $\omega_n = (2n+1)\pi/\beta$ is the Matsubara frequency.  We solve them numerically and re-insert into \eqref{euc_action_fermion} to obtain the free energy, hence the full thermodynamics(Sec.~\ref{thermodynamics}).  Consider the entropy $\mathtt S$.  A key feature of the SYK$_4$ solution is an extensive ($\propto N$) entropy\cite{maldacena_2016} in the $T\rightarrow 0$ limit, an extreme non-Fermi liquid feature.   This entropy must be removed over the narrow temperature window set by the the coherence energy $E_c$.  Consequently, we expect that $\mathtt S/N = \mathcal{S}(T/E_c)$ for $T,E_c \ll U_0$, where the universal function $\mathcal{S}(\mathcal{T}=0) = 0$ indicating no zero temperature entropy in a Fermi liquid, and $\mathcal S(\mathcal{T}\rightarrow \infty) = 0.4648\cdots$, recovering the zero temperature entropy of the SYK$_4$ model. The universal scaling collapse is confirmed by numerical solution, as shown in  Fig.~\ref{thermo}.  This implies also that the specific heat $NC = (T/E_c) \mathcal{S}'(T/E_c)$, and hence the low-temperature Sommerfeld coefficient
\begin{equation}
  \label{eq:4}
  \gamma\equiv\lim_{T\rightarrow 0}\frac{C}{T}=\frac{\mathcal{S}'(0)}{E_c}
\end{equation}
is {\em large} due to the smallness of $E_c$.  Specifically, compared with the Sommerfeld coefficient in the weak interaction limit $t_0\gg U_0$, which is of order $t_0^{-1}$, there is an ``effective mass enhancement'' of $m^*/m \sim t_0/E_c \sim U_0/t_0$.  Thus the low temperature state is a {\em heavy Fermi liquid}.
 
 To establish that the low temperature state is truly a strongly renormalized Fermi liquid with large Fermi liquid parameters, we compute the compressibility, $NK = \left. \partial \mathcal N/\partial \mu\right|_T$.  Because the compressibility has a smooth low temperature limit in SYK$_4$ model, we expect that $K$ is only weakly perturbed by small $t_0$.  For $t_0 \ll U_0$, we indeed have $K \approx \left. K\right|_{t_0=0} = c/U_0$ with the constant $c\approx 1.04$ regardless of $T/E_c$.  For free fermions, the compressibility and Sommerfeld coefficient are both proportional to the single-particle density of states (DOS), and in particular $\gamma/K = \pi^2/3$ for free fermions.  Here we find $\gamma/K = (\mathcal{S}'(0)/c) U_0/E_c \sim (U_0/t_0)^2 \gg 1$.  This can only be reconciled with Fermi liquid theory by introducing a large Landau interaction parameter.  In Fermi liquid theory, one introduces the interaction $f_{ab}$ via $\delta\varepsilon_a=\sum_b f_{ab}\delta n_b$, where $a,b$ label quasiparticle states.   For a diffusive disordered Fermi liquid, we take $f_{ab} = F/g(0)$, where $g(0)$ is the quasi-particle DOS, and $F$ is the dimensionless Fermi liquid interaction parameter.  The standard result of Fermi liquid theory(Supplementary Information), is that $\gamma$ is unaffected by $F$ but $K$ is renormalized, leading to $\gamma/K = \frac{\pi^2}{3} (1+F)$.   We see that $F \sim (U_0/t_0)^2 \gg 1$, so that the Fermi liquid is extremely strongly interacting.  Comparing to the effective mass, one has $F \sim (m^*/m)^2$.
 
\begin{figure}[htbp]
 \begin{center}
 \includegraphics[width=0.4\textwidth]{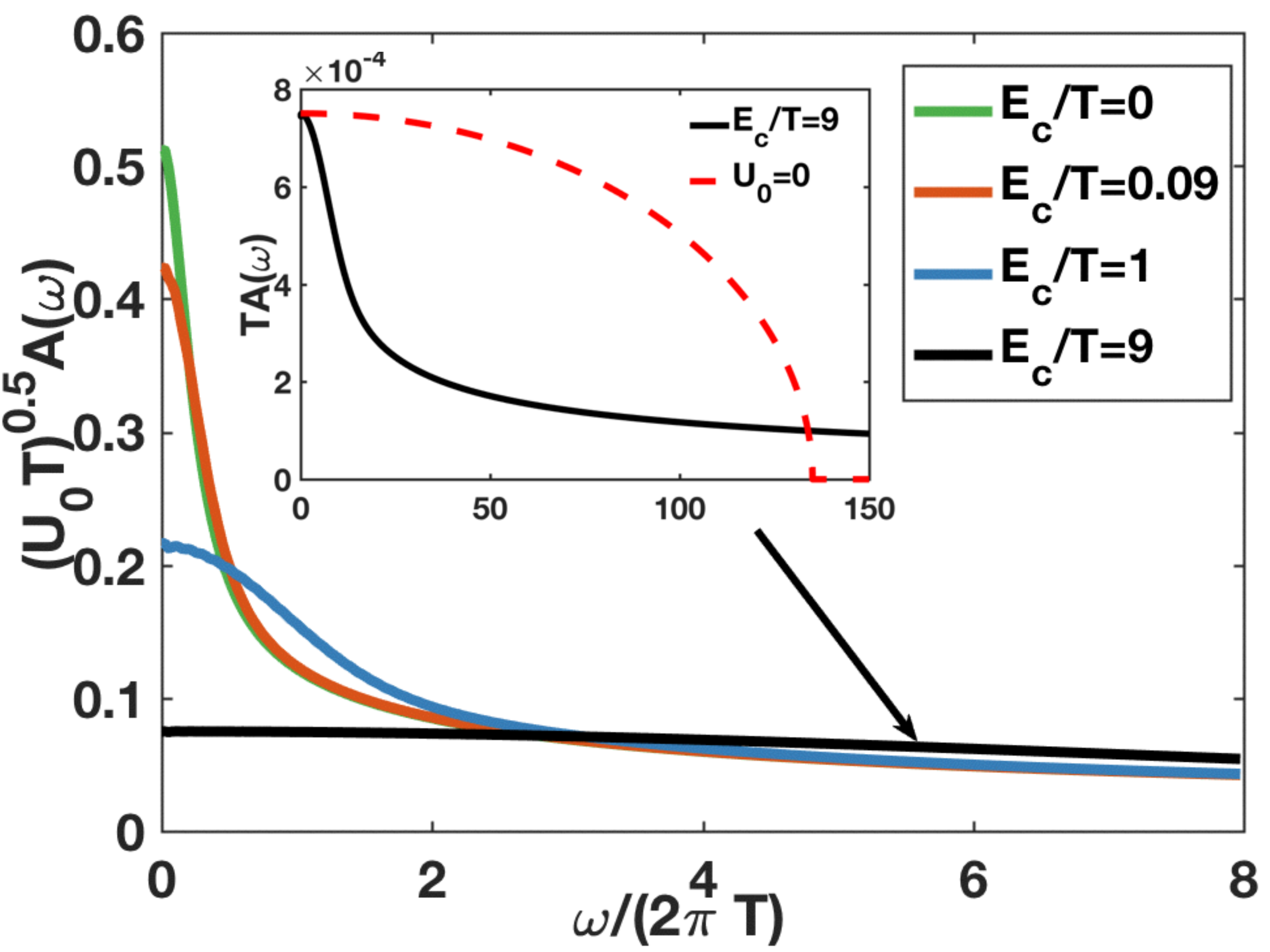}
 \caption{ The spectral weight $A(\omega)$ at fixed $U_0/T=10^4,\mu=0,z=2$ for $E_c/T=0,0.09,1,9$, corresponding a crossover from SYK$_4$ limit to the ``heavy Fermi liquid" regime.  Inset shows the comparison of green's function for $T/E_c=9$ with free fermion limit result.
 }    \label{green}
 \end{center}
 \end{figure}

\emph{Real time formulation-}  While imaginary time formulation is adequate for thermodynamics, it encounters difficulties in addressing transport due to difficulty of analytic continuation to zero {\em real} frequency in the presence of the emergent low energy scale $E_c$.  Instead we reformulate the problem in real time using Keldysh path integral. 
The Keldysh formalism calculates the partition function $Z=\frac{\text{Tr}[\rho U]}{\text{Tr}[\rho]}$ with $\rho=e^{-\beta(\mathcal H-\mu\mathcal N)}$ and $U$ the \emph {identity} evolution operator $U=e^{-i(\mathcal H-\mu\mathcal N)(t_0-t_f)}e^{-i(\mathcal H-\mu\mathcal N)(t_f-t_0)}$ describing evolving forward from $t_0\rightarrow t_f$ (with Keldysh label $+$) and backward (Keldysh label $-$) identically. 
Paralleling the imaginary-time development, we introduce collective variables $G_{x,ss'}(t,t')=\frac{-i}{N}\sum_i c^s_{ixt}\bar c^{s'}_{ixt'}$ and  $\Sigma_{x,ss'}$ with $s,s'=\pm$ labeling Keldysh contour, and integrate out the fermionic fields to obtain $\bar Z=\int [dG][d\Sigma] e^{iNS_K}$ (Sec.~\ref{keldysh} in Methods), with the Keldysh action
\begin{widetext}
\begin{eqnarray}
\label{keldysh_action}
iS_K&=& \sum_x\ln\det[\sigma^z (i\partial _t +\mu)\delta(t-t')- \Sigma_{x}(t,t')]
-\sum_{ss'}\int_{t_0}^{t_f} dt dt' \Big[\sum_x\frac{U_0^2}{4}ss' G_{x,ss'}(t,t')^2G_{x,s's}(t',t)^2-\sum_x\Sigma_{x,ss'}(t,t') G_{x,s's}(t',t)\nonumber\\&
+&\sum_{\bra x'x\ket} t_0^2ss' G_{x,ss'}(t,t')G_{x',s's}(t',t)\Big]
\end{eqnarray}
\end{widetext}
where $\Sigma_{x}$ in the determinant is to be understood as the matrix $[\Sigma_{x,ss'}]$ and $\sigma^z$ acts in Keldysh space. 
 We obtain the numerical solution to the Green's functions(Sec.~\ref{numerics_green}) by solving for the saddle point of $S_K$.   We plot in Fig.~\ref{green} the spectral weight $A(\omega)\equiv \frac{-1}{\pi}\text{Im}\,G_R(\omega)$ ($G_R$ is retarded Green function) at fixed $U_0 /T =10^4$ for $E_c/T=0,0.09,1,9$, which illustrates the crossover between the SYK$_4$ and Fermi liquid regimes.  For $\omega \gg E_c$, we observe the quantum critical form of the SYK$_4$ model, which displays $\omega/T$ scaling,  evident in the figure from the collapse onto a single curve at large $\omega/T$.  At low frequency, the SYK$_4$ model has $A(\omega \ll T) \sim 1/\sqrt{U_0 T}$, whose divergence as $T\rightarrow 0$ is cut-off when $T \lesssim E_c$.   This is seen in the reduction of the peak height in Fig.~\ref{green}, $\sqrt{U_0 T}A(\omega=0)$, with increasing $E_c/T$.  On a larger frequency scale (inset), the narrow ``coherence peak'', associated with the small spectral weight of heavy quasiparticles, is clearly visible.


 We now turn to transport, and for simplicity focus on particle-hole symmetric case hereafter.  The strategy is to obtain electrical and heat conductivities from the fluctuations of charge and energy, respectively, using the Einstein relations.  We first consider charge, and study the low-energy $U(1)$ phase fluctuation $\varphi(x,t)$, which is the conjugate variable to particle number density $\mathcal N(x,t)$, around the saddle point of the action $S_K$.   Allowing for phase fluctuations around the saddle point solution amounts to taking
\begin{eqnarray}
\label{phi_var}
G_{x,ss'}(t,t')\rightarrow G_{x,ss'}(t-t')e^{-i(\varphi_s(x,t)-\varphi_{s'}(x,t'))}\nonumber\\
\Sigma_{x,ss'}(t,t')\rightarrow\Sigma_{x,ss'}(t-t')e^{-i(\varphi_s(x,t)-\varphi_{s'}(x,t'))},
\end{eqnarray}
where $G_{x,ss'}(t-t')$ and $\Sigma_{x,ss'}(t-t')$ are the saddle point solutions.
Expanding \eqref{keldysh_action} to quadratic order in $\varphi_s$, $S_K = S_K^{sp} + S_\varphi$, yields the lowest order effective action for the $U(1)$ fluctuations. This is most conveniently expressed in terms of the  ``classical'' and  ``quantum" components of the phase fluctuations, defined as $\varphi_{c/q}=(\varphi_+\pm \varphi_-)/2$ and in Fourier space:
\begin{eqnarray}
\label{phi_action}
iS_{\varphi}&=&\sum_{\bf p}\!\int_{t_0}^{t_f}\!dtdt' \big[\Lambda_1(t-t')\partial_t \varphi_{c,{\bf p}} (t)\partial_t \varphi_{q,-{\bf p}}(t')
\nonumber\\ && -\upsilon ({\bf p})\Lambda_2(t-t')\varphi_{c,{\bf p}}(t)\varphi_{q,-{\bf p}}(t')\big]+\cdots.
\end{eqnarray}
Here the first term arises from the $\ln \det [\cdot]$ and the second from the hopping ($t_0^2$) term in \eqref{keldysh_action}.  The function
$\upsilon ({\bf p})$ encodes the band structure for the two-fermion hopping term, dependent on lattice details, and the ellipses represent $O(\varphi_q^2)$ terms which do not contribute to the density correlations (and are omitted hereafter --see Sec.~\ref{keldysh} for reasons).  The coefficients $\Lambda_1(t)$ and $\Lambda_2(t)$ are expressed in terms of saddle point Green's functions in Sec.~\ref{keldysh}.   
We remark here that any further approximations, e.g., conformal invariance, are not assumed to arrive at action \eqref{phi_action}, and hence this derivation applies in \emph{all} regimes.

In the low frequency limit, the Fourier transforms of  $\Lambda_1(t)$,$\Lambda_2(t)$ behave as $\Lambda_1(\omega)\approx -2iK$ and $\Lambda_2(\omega)\approx 2KD_{\varphi} \omega$, which {\em defines} the positive real parameters $K$ and $D_\varphi$.  At small momentum, for an isotropic Bravais lattice, $\upsilon ({\bf p})=p^2$ (with unit lattice spacing), and the phase action becomes
\begin{eqnarray}
\label{phi_diffusive}
iS_{\varphi}&=& -2K\sum_p\int_{-\infty}^{+\infty} \!\!\! d\omega\,  \varphi_{c,\omega}(i\omega^2-D_{\varphi}p^2\omega)\varphi_{q,-\omega} .
\end{eqnarray}
The density-density correlator is expressed as
\begin{eqnarray}
\label{re_NN}
D_{Rn}(x,\!t;x'\!,\!t')&\equiv& i\theta(t-t')\bra[\mathcal N(x,\!t),\mathcal N(x'\!,\!t')]\ket\nonumber\\
&=&\frac{i}{2}\bra \mathcal N_c(x,\!t)\mathcal N_q(x'\!,\!t')\ket,
\end{eqnarray}
where  $\mathcal N_s\equiv\frac{N\delta S_\varphi}{\delta \dot\varphi_s}$, $\mathcal N_{c/q}=\mathcal N_+\pm \mathcal N_-$(keeping momentum-independent components- See Sec.\ref{keldysh}). Adding a contact term to ensure that $\lim_{p\rightarrow 0} D_{Rn}(p,\omega\neq 0)=0$\cite{policastro_2002}, the action~\eqref{phi_diffusive} yields the diffusive form \cite{kadanoff_martin}
\begin{equation}
\label{dr_form}
D_{Rn}(p,\omega)=\frac{-iNK\omega}{i\omega-D_\varphi p^2}+NK=\frac{-NKD_{\varphi}p^2}{i\omega-D_{\varphi}p^2}.
\end{equation}
From this we identify $N K$ and $D_{\varphi}$ as the compressibility and charge diffusion constant, respectively.  
The electric conductivity is given by Einstein relation $\sigma\equiv 1/\rho=NKD_{\varphi}$, or, restoring all units,$\sigma = NK D_\varphi \frac{e^2}{\hbar}a^{2-d}$($a$ is lattice spacing).  Note the proportionality to $N$: in the standard non-linear sigma model formulation, the dimensionless conductance is large,  suppressing localization effects.  This occurs because both $U$ and $t$ interactions scatter between all orbitals,  destroying interference from closed loops.

The analysis of energy transport proceeds similarly.  Since energy is the generator of time translations,  one considers the time-reparametrization (TRP) modes induced by $t_s\rightarrow t_s+\epsilon_s(t)$ and defines $\epsilon_{c/q}=\frac{1}{2}(\epsilon_+\pm\epsilon_-)$. 
The effective action for TRP modes to the lowest-order in $p,\omega$ reads (Sec.~\ref{keldysh}) 
\begin{equation}
\label{ep_action}
iS_{\epsilon}=\sum_p\int_{-\infty}^{+\infty}\!\!\! d\omega\,  \epsilon_{c,\omega} ( 2i\gamma\omega^2T^2-p^2\Lambda_3(\omega))\epsilon_{q,-\omega}+\cdots, 
\end{equation}
where the ellipses has the same meaning as in \eqref{phi_diffusive}.  At low frequency, the correlation function integral, given in Sec.~\ref{keldysh}, behaves as $\Lambda_3(\omega)\approx 2\gamma D_{\epsilon}T^2 \omega$, which defines the energy diffusion constant $D_{\epsilon}$.  This identification is seen from the correlator for energy density modes $\varepsilon_{c/q} \equiv \frac{iN\delta S_\epsilon}{\delta \dot\epsilon_{c/q}} $, 
\begin{eqnarray}
D_{R\varepsilon}(p,\omega)=\frac{i}{2}\bra \varepsilon_c\varepsilon_q\ket_{p,\omega}=\frac{-NT^2\gamma D_{\epsilon}p^2}{i\omega-D_{\epsilon}p^2},
\end{eqnarray}
where we add a contact term to ensure conservation of energy at $p=0$.  The thermal conductivity reads $\kappa=NT\gamma D_{\epsilon}$ ($k_B=1$) --like $\sigma$, is $O(N)$.
 
 \begin{figure}[htbp]
 \begin{center}
 \subfigure[]{ 
\label{con1} 
\includegraphics[width=0.4\textwidth]{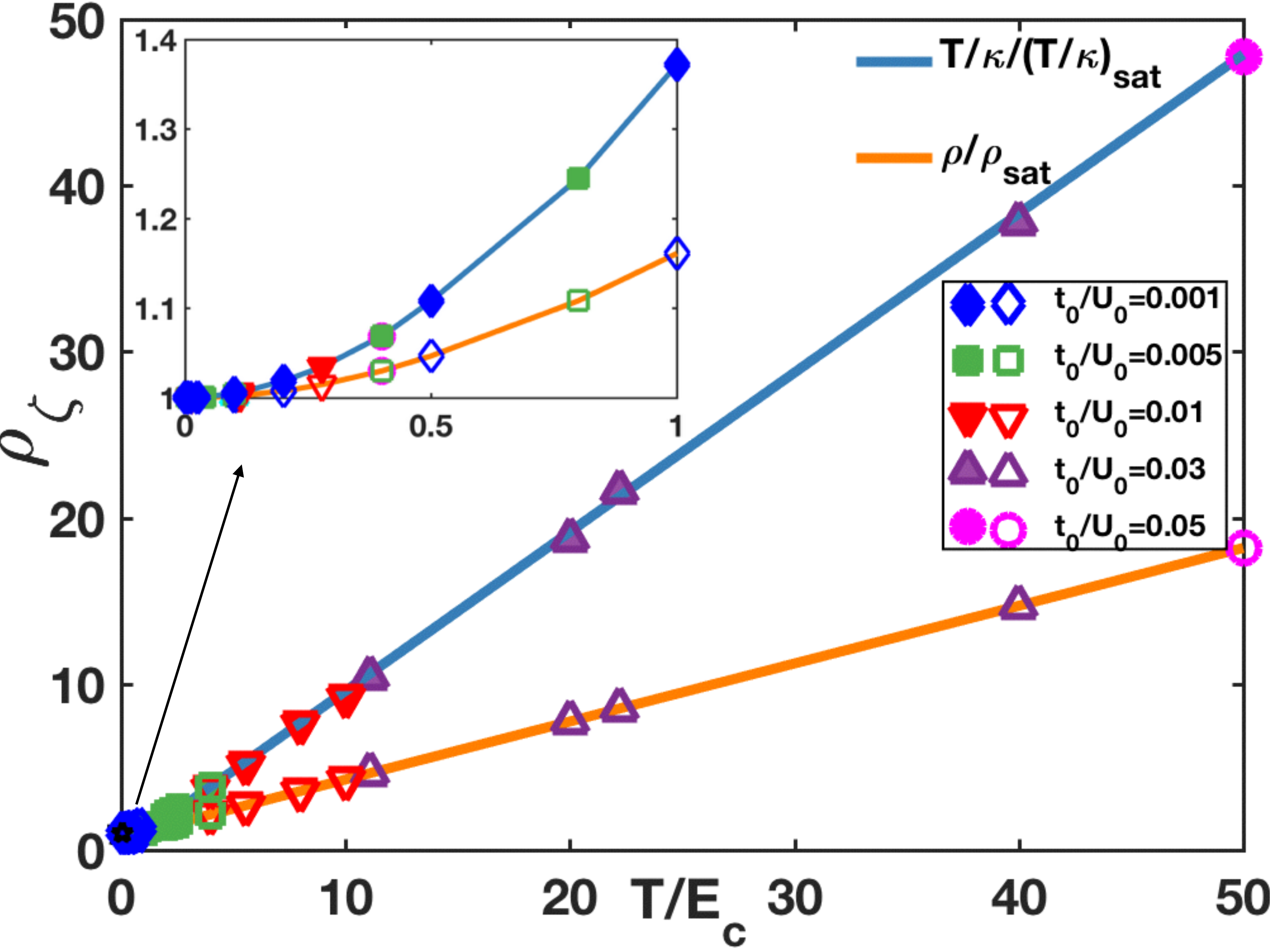}} 
\subfigure[]{ 
\label{con2} 
\includegraphics[width=0.4\textwidth]{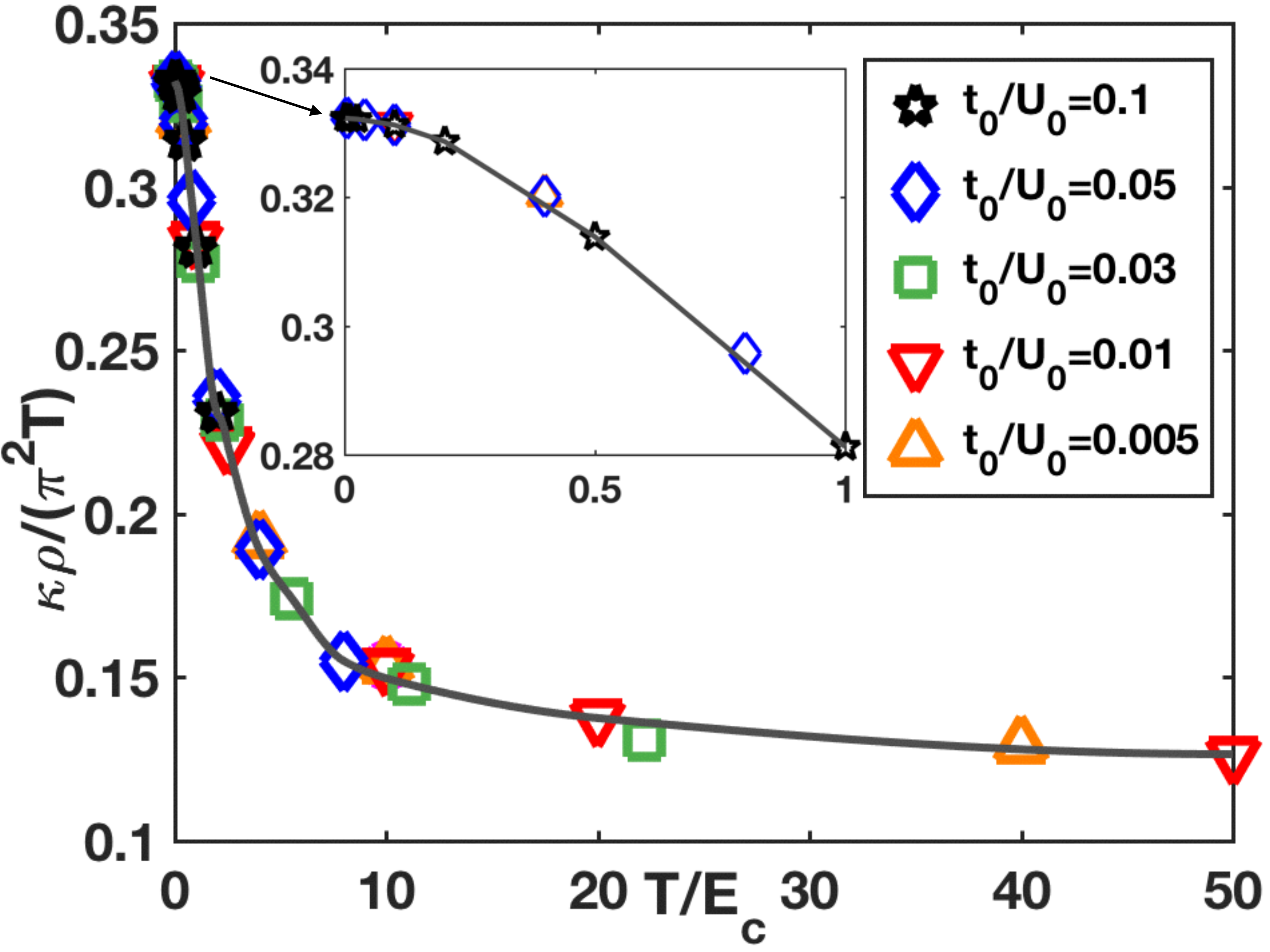} }
\caption{(a): For $t_0, T\ll U_0$, $\rho_{\varphi/\varepsilon}$ ``collapse'' to $R_{\varphi/\varepsilon}(\frac{T}{E_c})/N$. (b): The Lorentz ratio $\frac{\kappa\rho}{T}$ reaches two constants $\frac{\pi^2}{3},\frac{\pi^2}{8}$, in the two regimes. The solid curves are guides to the eyes.} \label{con}
 \end{center}
 \end{figure}
 
\emph{Scaling collapse, Kadowaki-Woods and  Lorentz ratios} -- Electric/thermal conductivities are obtained from  $\lim_{\omega\rightarrow 0} \Lambda_{2/3}(\omega)/\omega$, expressed as integrals of real-time correlation functions, and can be evaluated numerically for any $T,t_0,U_0$.  Introducing generalized resistivities, $\rho_\varphi=\rho$, $\rho_\varepsilon = T/\kappa$, we find remarkably that for $t_0, T\ll U_0$, they collapse to universal functions of one variable,
\begin{equation}
  \label{eq:3}
  \rho_\zeta(t_0,T \ll U_0) = \frac{1}{N}R_\zeta( \tfrac{T}{E_c}) \qquad \zeta\in \{\varphi,\varepsilon\},
\end{equation}
where $R_\varphi(\mathcal{T})$, $R_\varepsilon({\mathcal T})$ are dimensionless universal functions.   This scaling collapse is verified by direct numerical calculations shown in Fig.~\ref{con}a.    From the scaling form~\eqref{eq:3}, we see the {\em low temperature} resistivity obeys the usual Fermi liquid form
\begin{equation}
  \label{eq:2}
  \rho_\zeta(T \ll E_c) \approx \rho_\zeta(0) + A_\zeta T^2,
\end{equation}
where the temperature coefficient of resistivity $A_\zeta = \frac{R''_\zeta(0)}{2NE_c^2}$ is large due to small coherence scale in denominator, characteristic of a strongly correlated Fermi liquid.  Famously, the Kadowaki-Woods ratio, $A_\varphi/(N\gamma)^2$, is approximately system-independent for a wide range of correlated materials\cite{kadowaki1986universal,rice1968electron}. We find here $\frac{A_\varphi}{(N\gamma)^2} = \frac{R''_\varphi(0)}{2[\mathcal{S}'(0)]^2N^3}$ is independent of $t_0$ and $U_0$!

Turning now to the incoherent metal regime, in limit of large arguments, $\mathcal{T} \gg 1$, the generalized resistivities vary linearly with temperature: $R_\zeta(\mathcal{T}) \sim c_\zeta\, \mathcal{T}$.  We  analytically obtain $c_\varphi = \frac{2}{\sqrt{\pi}}$ and $c_\varepsilon = \frac{16}{\pi^{5/2}}$ (Supplementary Information), implying that the Lorenz number, characterizing the Wiedemann-Franz law, takes the unusual value $L=\frac{\kappa}{\sigma T}\rightarrow \frac{\pi^2}{8}$ for $E_c \ll T \ll U_0$.  More generally, the scaling form~\eqref{eq:3} implies that $L$ is a universal function of $T/E_c$, verified numerically as shown in Fig.~\ref{con}b.  The Lorenz number increases with lower temperature, saturating at $T\ll E_c$ to the Fermi liquid value $\pi^2/3$.  
 
\emph{Conclusion} -- We have shown that the SYK model provides a soluble source of strong local interactions which, when coupled into a higher-dimensional lattice by ordinary but random electron hopping, reproduces a remarkable number of features of strongly correlated metals, including heavy quasiparticles with small spectral weight, a largely system-independent Kadowaki-Woods ratio, $T$-linear high temperature resistivity, and an anomalous Lorenz number in the incoherent regime.  The remarkable success of this simple soluble model suggests exciting prospects for extending the treatment to more realistic systems, and to shed light on the physical content of various numerical results from dynamical mean field theory\cite{georges_1996}, which shares significant mathematical similarity to basic equations of this work.  

 \emph {Acknowledgements:}  X.-Y.~S thanks Wenbo Fu, Subir Sachdev and in particular Yingfei Gu for helpful discussions and lectures.  Work by X.-Y.~S was supported by the ARO, Grant No.~W911-NF-14-1-0379 and the National Innovation Training Program at PKU.  Work by C.-M.~J. was supported by the Gordon and Betty Moore Foundation(Grant~$4034$).  Work by L. B. was supported by the DOE, Office of Science, Basic Energy Sciences under award number DE-FG02-08ER46524.  The research benefitted from facilities of the KITP, by grant No. NSF PHY-1125915 and Center for Scientific Computing from the CNSI, MRL under grant NSF MRSEC (DMR-1121053) and NSF CNS-0960316. 
 
 \emph{Author contribution:} All authors participated in theoretical construction/derivations. X.-Y~S performed numerical calculation under supervision of L.B. and C.-M~J.
 
 The authors declare no competing financial interests.

 \appendix
 \renewcommand{\appendixname}{}

 \begin{widetext}
{\bf \large Methods:}
\section{Symmetries and Green's function formulated in Euclidean space}
 \label{euclidean}

 \subsection{Euclidean action and symmetries}
 The action in Eq.~\eqref{eq:1} and its counterpart for collective variables, Eq.~\eqref{euc_action_fermion}, enjoy both a $U(1)$  symmetry corresponding to charge conservation, as well as time-translation symmetry corresponding to energy conservation.  The latter is elevated to a full time-reparametrization(TRP) symmetry in the ``conformal limit'' in which we neglect both the time-derivative and the hopping term.  Together these act as follows:
\begin{eqnarray}
\label{time_rep}
f(\tau)\in\text{Diff}(S^1)\quad\tau\rightarrow f(\tau)&\quad& c(\tau)\rightarrow e^{i\varphi(\tau)}f'(\tau)^{\frac{1}{4}}c(f(\tau))\quad \bar c(\tau)\rightarrow e^{-i\varphi(\tau)}f'(\tau)^{\frac{1}{4}}\bar c(f(\tau))\nonumber\\
G(\tau,\tau')\rightarrow e^{i(\varphi(\tau)-\varphi(\tau'))}f'(\tau)^{\frac{1}{4}}f'(\tau')^{\frac{1}{4}} G(f(\tau),f(\tau'))&\quad& \Sigma(\tau,\tau')\rightarrow e^{i(\varphi(\tau)-\varphi(\tau'))}f'(\tau)^{\frac{3}{4}}f'(\tau')^{\frac{3}{4}} \Sigma(f(\tau),f(\tau')).
\end{eqnarray}
 
 \subsection{Saddle-point solution}
Differentiating the action w.r.t. $G,\Sigma$, we have for the saddle point condition as
\begin{eqnarray}
\label{SD}
 &&\frac{\delta S}{\delta G}: -U_0^2 G_x(\tau,\tau')^{2}G_x(\tau,\tau')+t_0^2\sum_{\langle x',x\rangle} G_{x'}(\tau,\tau')=\Sigma_x(\tau,\tau'),\nonumber\\
 &&\frac{\delta S}{\delta \Sigma}: \frac{1}{i\omega+\mu-\Sigma_x(i\omega)}=G_x(i\omega).
 \end{eqnarray}
 
 The solution to the Schwinger-Dyson equation is site-independent. In the conformal limit, the solution of two-point function is given in Ref.~\onlinecite{davidson_2016}:
 \begin{eqnarray}
 \label{saddle}
 G_{\text{sad}}(\tau)=\frac{-b e^{-\mathcal E \phi}}{\sqrt {\sin(\frac{\phi}{2})}} \quad (\tau>0),\quad G_{\text{sad}}(\tau)=\frac{b e^{-2\pi\mathcal E-\mathcal E\phi}}{\sqrt {\sin(\frac{-\phi}{2})}} \quad (\tau<0),
  \end{eqnarray}
with $\phi\equiv\frac{2\pi \tau}{\beta}$ and $b=\frac{\pi^{\frac{1}{4}}}{\sqrt {2\beta U_0}}$, and $2\pi\mathcal E$ is related to the spectral asymmetry as discussed in Refs \onlinecite{davidson_2016,sachdev_2015}. In particular, $\mathcal E = 0$ in the presence of particle hole symmetry ($\mu = 0$). 
The Fourier transformation of the two-point function reads $ G_{\text{sad}}(i\omega_n)=\frac{-i\sqrt\beta C}{\sqrt{2U_0 \pi^{\frac{1}{2}}}}\frac{\Gamma(\frac{3}{4}+n+i\mathcal E)}{\Gamma(\frac{5}{4}+n+i\mathcal E)}$
  where $\omega_n=\frac{2\pi(n+1/2)}{\beta}, C=\frac{(1+i)(1+ie^{-2\mathcal E \pi})}{2i}$.

 \section{Effective Keldysh action for $\varphi, \epsilon$ fields}
 \label{keldysh}
 \subsection{Keldysh action}
In the Keldysh approach discussed in the main text, the partition function is written using coherent states as $Z=\int [d\bar c][d c]e^{iS}$, with two sets of Grassmann variables on the forward and backward time contours labeled by $s=+1(+),-1(-)$, respectively:
 \begin{eqnarray}
&& S= \sum_s\left\{\int_{t_0}^{t_f} \sum_{x,i} \bar c_{s,i,x} is\partial_t c_{s,i,x}-s\int_{t_0}^{t_f}dt \Big[\sum_{x,i<j,k<l} U_{ijkl,x}\bar c_{s,ix}\bar c_{s,jx} c_{s,kx}c_{s,lx}+\sum_{\langle xx'\rangle,i,j}(t_{ij,x}\bar c_{s,i,x} c_{s,j,x'}+h.c.)-\mu \sum_{x,i}\bar c_{s,i,x} c_{s,i,x}\Big]\right\}.
 \end{eqnarray}
 We take Gaussian distributed disorder,  $P(U)=\sqrt{\frac{N^3}{2\pi U_0^2}}e^{-\frac{N^3}{2U_0^2}|U|^2}, P(t)=\sqrt{\frac{N}{\pi t_0^2}}e^{-\frac{N}{t_0^2}|t|^2}$. The disorder-averaged partition function reads $\bar Z=\bra Z\ket_{dis}=\int [dc][d\bar c]e^{iS}$, with
\begin{eqnarray}
 \label{disorder_Z}
&&S=\int_{t_0}^{t_f} dt i \sum_{s,a,x}s\bar c_{s,a,x} \partial_t c_{s,a,x}+
\left \{
\sum_{x,i<j,k<l}\frac{i2U_0^2}{N^3} \Bigg|\sum_{s}s\int _{t_0}^{t_f} dt \bar c_{s,ix} (t)\bar c_{s,jx}(t)c_{s,kx}(t)c_{s,lx}(t) \Bigg|^2
\right.
\nonumber\\
&&  
\left.
+ \sum_{\langle x,x'\rangle ,i,j}\frac{ it_0^2}{ N}  \Bigg|\sum_{s} s\int_{t_0}^{t_f} dt\bar c_{s,i,x}(t)c_{s,j,x'}\Bigg|^2+\mu\int_{t_0}^{t_f} dt \sum_{s,x,i}s\bar c_{s,i,x}(t)c_{s,i,x}(t)
\right \} .
 \end{eqnarray}

We introduce a composite field $
   G_{x,ss'}(t,t')=\frac{-i}{N}\sum_a  c_{s,a,x}(t) \bar c_{s',a,x}(t')$ together with a Lagrange multiplier $\Sigma _{s,s'}(t,t')$ imposing this identity, which leads to
\begin{eqnarray}
&& S=\int_{t_0}^{t_f} dt \sum_{s,a,x}s\bar c_{s,a,x}(t)(i\partial_t+\mu) c_{s,a,x}(t)+\sum_{s,s'}\int _{t_0}^{t_f} dt dt' \Bigg\{\sum_x\frac{iU_0^2N}{4}ss' G_{x,ss'}(t,t')^2G_{x,s's}(t',t)^2\nonumber\\
&&+ \sum_{\langle x,x'\rangle} it_0^2 N ss' G_{x,ss'}(t,t')G_{x',s's}(t',t)-iN\sum_{x} \Sigma_{x,ss'}(t,t')\Big[G_{x,s's}(t',t)-\frac{i}{N}\sum_a \bar c_{s,a,x}(t) c_{s',a,x}(t')\Big]\Bigg\}.
\end{eqnarray}
Integrating out fermion fields,  we are left with Eq.~\eqref{keldysh_action} of the main text.

\subsection{Keldysh rotation and real time Green's function}
The saddle point conditions for the action of Eq.~\eqref{keldysh_action} are, assuming space and time translational invariance (defining Fourier transforms $\Sigma(\omega)=\int dt\Sigma(t)e^{i\omega t} ,G(\omega)=\int dt G(t)e^{i\omega t} $)
\begin{eqnarray}
\label{SDk}
&&(\sigma^z(\omega+\mu)-\Sigma(\omega))^{-1}=  G_{ss'}(\omega),\nonumber\\
&&ss' U_0^2  G_{ss'}(t)^2 G_{s's}(-t)+zt_0^2ss'  G_{ss'}(t)=\Sigma_{ss'}(t).
\end{eqnarray}
where $z$ is the coordination number of the lattice in consideration. A standard rotation relates $G_{ss'}$ to more conventional Green's functions (see Supplementary Information).  With $L=\frac{1}{\sqrt{2}}\left( \begin{array}{cc}
1 & -1 \\
1 & 1
\end{array} \right)$, we can write
\begin{eqnarray}
\label{keldysh_g}
\left (\begin{array} {cc} G_R & G_K \\ 0 &G_A\end{array} \right )= L \sigma^z  G L^\dagger
=\frac{1}{2} \left(\begin{array} {cc}  G_{++}- G_{+-}+ G_{-+}- G_{--} & G_{++}+ G_{+-}+ G_{-+}+ G_{--} \\ 0 & G_{++}+ G_{+-}- G_{-+}- G_{--}\end{array}\right),
\end{eqnarray}
where $G_R,G_A$ are  the usual retarded and advanced green's functions, and the Keldysh green's function $G_K(t_1,t_2)=-i \bra [c(t_1),c^\dagger(t_2)]\ket$. 

On inspection, one sees that \eqref{SDk} do not involve temperature $T$ at all.  This is a feature of the Keldysh technique since temperature enters only through the initial density matrix at $t_0= -\infty$.   It implies that there are distinct saddle point solutions corresponding to each temperature, as well as possible non-equilibrium ones.  For our purpose, we enforce equilibrium at temperature $T$ through the fluctuation-dissipation relation, $G_K(\omega) =2i\tanh(\frac{\beta\omega}{2})\text{Im}[G_R(\omega)]$ (Supplementary Information).  The numerical implementation of this is discussed in Sec.~\ref{numerics_green}.   

In the conformal limit, we can ``cheat'' and obtain the solutions by analytic continuation of the imaginary time result.  We find (see Supplementary Information for details):
\begin{eqnarray}
\label{G_re}
&&G_R(t)=-i\sqrt 2 b\frac{\theta(t)}{\sqrt{\sinh [\frac{\pi |t|}{\beta}]}},\qquad
G_A(t)=i\sqrt 2 b\frac{\theta(-t)}{\sqrt{\sinh [\frac{\pi |t|}{\beta}]}},\qquad G_K(t)=- \sqrt 2 b\frac{sgn(t)}{\sqrt{\sinh [\frac{\pi |t|}{\beta}]}}.
\end{eqnarray}
The Fourier transform reads
\begin{eqnarray}
\label{G_re_f}
&&G_R(\omega)=\frac{-ib\beta}{\sqrt {\pi}}\frac{\Gamma(1/4-\frac{i\beta\omega}{2\pi})}{\Gamma(3/4-\frac{i\beta\omega}{2\pi})},\qquad G_A(\omega)=G_R^*(\omega).
\end{eqnarray}


\subsection{Effective action for phase fluctuations $\varphi$ and density-density correlations}
Before proceeding further, we remark that when $\mu=0$, the fluctuations for $U(1)$ and TRP modes associated $\varphi,\epsilon$ fields are decoupled to quadratic order and we could obtain charge/energy transport from the action for $\varphi,\epsilon$ separately. We therefore assume zero chemical potential in the following.  

Including the phase fluctuations around the saddle point following Eq.~\eqref{phi_var}, we obtain from Eq.~\eqref{keldysh_action}
\begin{eqnarray}
\label{kel_action}
&&\frac{iS}{N}=\sum_x\text{Tr}\{\ln[(\sigma^z (i\partial_t+\mu)+\sigma^z \partial_t \varphi)\delta(t-t')-\Sigma(t,t')]\}-t_0^2\sum_{\bra xx'\ket,ss'}\!\int_{-\infty}^{+\infty} dt dt'ss' G_{x,ss'}(t,t') G_{x',s's}(t',t)e^{-i(\nabla_x \varphi_{s'}(t')\cdot {\bf l}_{xx'}-\nabla_x\varphi_s(t)\cdot{\bf l}_{xx'})}\nonumber\\
&&=\text{const}+\frac{-1}{2}\sum_x\text{Tr}[G\sigma^z \partial_t \varphi G\sigma^z \partial_t \varphi]+\frac{t_0^2}{2}\sum_{s,s',p}\upsilon({\bf p})ss'\int _{-\infty}^{+\infty}dt dt' G_{ss'}(t,t')G_{s's}(t',t)[\varphi_{s'}(t')^2+\varphi_{s}(t)^2-2\varphi_s(t)\varphi_{s'}(t')],
\end{eqnarray}
where $\upsilon ({\bf p})=\frac{1}{2}\sum_{x' \text{adjacent to}x}(\bf p\cdot \bf l_{xx'})^2$, where one sums over all neighboring sites $x'$ of some site $x$, and $\bf l_{xx'}$ denotes the displacement from site $x$ to site $x'$, and $\varphi=\text{diag}(\varphi_+(t),\varphi_-(t))$.   We absorbed the phase of $\Sigma$ in the $\text{Tr}\ln$ by an associated gauge transformation of the Grassmann fields, i.e. $c_s(t)\rightarrow c_s e^{i\varphi_s(t)},\bar c_s(t)\rightarrow\bar c_s e^{-i\varphi_s(t)}$, resulting in the $\sigma^z \partial_t \varphi$ term. After some algebra, we get the effective action  ($\bar Z=\int [d \varphi]e^{iNS_\varphi}$)
\begin{eqnarray}
\label{phi_action1}
iS_{\varphi}&=&\sum_p\!\int_{-\infty}^{+\infty} dt dt' [\Lambda_1(t-t')\partial_t \varphi_c(t)\partial_t \varphi_q(t')+\frac{1}{2}\Lambda_4(t-t') \partial_t \varphi_q(t)\partial_t \varphi_q(t')
-\upsilon ({\bf p})\Lambda_2(t-t')\varphi_c(t)\varphi_q(t')-2t_0^2 \Lambda_4(t-t')\varphi_q(t)\varphi_q(t')]\nonumber\\
&=&\sum_p\int_{-\infty}^{+\infty} d\omega \varphi_{c,\omega}[\Lambda_1(-\omega)\omega^2-\upsilon ({\bf p}) \Lambda_2(-\omega)]\varphi_{q,-\omega}+\Lambda_4(\omega)(\frac{1}{2}\omega^2-2t_0^2)|\varphi_{q.\omega}|^2
\end{eqnarray}
where
\begin{eqnarray}
\Lambda_1(t)&=&2i\text{Im}[G_K(t)G_R(-t)],\quad \Lambda_2(t)=2it_0^2[2\text{Im}[G_K(-t)G_R(t)]
-\delta(t)\int_{-\infty}^{+\infty} dt'2\text{Im}[G_K(t')G_R(-t')]],\nonumber\\
\Lambda_4(t)&=& |G_R(|t|)|^2-|G_K(t)|^2.
\end{eqnarray}
and $\Lambda_{1/2/4}(\omega)$ is Fourier transform of $\Lambda_{1/2/4}(t)$.  Hereafter we take $\upsilon({\bf p})=p^2$ (setting site spacing to $1$). 

The absence of any $O(\varphi_c^2)$ terms is a feature of the Keldysh technique, which implies that the retarded density correlator depends only upon the coefficient of $\varphi_c\varphi_q$ (see Supplementary Information).  Specifically, 
\begin{equation}
D_{Rn}(x,t)\equiv i\theta(t)\bra[\mathcal N(x,t),\mathcal N(0,0)]\ket =\frac{i}{2}\bra \mathcal N_c(x,t)\mathcal N_q(0,0)\ket,
\end{equation}
where $\mathcal N_s\equiv \frac{N\delta S_\varphi}{\delta \dot\varphi_s}, \mathcal N_{c/q}=\frac{N\delta S_\varphi}{\delta \dot\varphi_{c/q}}=(\mathcal N_+\pm\mathcal N_-)$.   This justifies neglecting the $O(\varphi_q^2)$ term, resulting in Eq.~\eqref{phi_action} of the main text.  Then approximating $\Lambda_1(\omega)\approx \Lambda_1(0)(=-2iK), \Lambda_2(\omega)\approx 2KD_{\varphi} \omega$, one obtains Eq.~\eqref{phi_diffusive}.    Then from the first line of action \eqref{phi_action1} we have (keeping only momentum-independent components of the derivative \cite{davidson_2016} that yield diffusive behavior -- this can be justified by following carefully the definition of the local charge operator from the initial Hamiltonian)
\begin{eqnarray}
\mathcal N_{q,\omega}= -N\omega \varphi_{c,\omega}\Lambda_1(\omega)-\frac{1}{2}N\omega \varphi_{q,\omega}(\Lambda_4(\omega)+\Lambda_4(-\omega)),\qquad \mathcal N_{c,-\omega}=N\omega \varphi_{q,-\omega}\Lambda_1(\omega).
\end{eqnarray}
Hence from action \eqref{phi_action1} the correlator is (again approximating $\Lambda_1(\omega)\approx \Lambda_1(0)=-2iK$, omitting the vanishing $\bra \varphi_{q,\omega}\varphi_{q,-\omega}\ket$)
\begin{equation}
D_{Rn}(p,\omega)=i2N^2K^2\omega^2 \bra \varphi_q\varphi_c\ket_{p,\omega}+NK=\frac{-iNK\omega}{i\omega-D_\varphi p^2}+NK=\frac{-NKD_{\varphi}p^2}{i\omega-D_{\varphi}p^2}.
\end{equation}
 
Next we briefly analyze the behavior of the two kernel function $\Lambda_{1(2)}(t)$ in the action \eqref{phi_action}. 
$\Lambda_1(t)=2i\text{Im}[G_K(t)G_R(-t)]$
 is purely imaginary; hence $\Lambda_1(\omega)=-\Lambda_1(-\omega)^*$.  For small frequency, it is therefore justified to approximate $\Lambda_1(\omega)\approx\Lambda_1(0)=-2iK$, as done in the main text.   Physically this coefficient is to be identified with the compressibility\cite{davidson_2016}.  The direct calculation of the compressibility from real time correlations of the charge, however, has the usual textbook subtleties: because the total charge is a conserved quantity, the correlations vanish at zero momentum at any frequency.  Thus a proper calculation of compressibility requires {\em local} non-conservation of charge, which is accomplished by taking frequency to zero first and then momentum to zero.  The lack of momentum dependence in $\Lambda_1(t)$ indicates this subtlety is not very accessible here.  Therefore  to avoid these order of limits issues, we instead extract $K$ directly from thermodynamics, calculated using the imaginary time formulation. 

We have
\begin{eqnarray}
\label{s2}
&&\Lambda_2(\omega)=4t_0^2\int_0^{\infty}\! dt\,  i\, \text{Im}[G_K(-t)G_R(t)](e^{i\omega t}-1)\approx -4\omega t_0^2\int_0^{\infty} \! dt\,  t\,  G_K(-t)\text{Im}[G_R(t)],
 \end{eqnarray}
where we only track the terms with lowest (linear) order in frequency, and we used the reality of $G_K(-t)$.  Expressing the above Fourier transform as convolution of $G_{K/R}(\omega)$ and taking the $\omega\rightarrow 0$ limit, we have 
\begin{eqnarray}
&&\sigma/N=\lim_{\omega\rightarrow 0}\frac{\Lambda_2(\omega)}{2\omega}= \frac{it_0^2}{\pi}\int_{-\infty}^{\infty}d\omega_1 G_K(\omega_1)\frac{\partial{\text{Im}[G_R(\omega_1)]}}{\partial \omega}=\pi t_0^2\beta\int_{-\infty}^{\infty} d\omega \frac{A(\omega)^2}{\cosh(\frac{\beta\omega}{2})^2},
\end{eqnarray}
where we used the fluctuation-dissipation relation $G_K=2i\tanh(\frac{\beta\omega}{2})\text{Im}[G_R(\omega)]$ (supplementary information) and performed integration by parts to arrive at the second line. One readily identify the above expression with the DC limit of conductivity by given by Kubo formula \cite{georges_1996} $\text{Re}\sigma(\omega)\propto t_0^2\int d\nu \frac{n_F(\nu+\omega)-n_F(\nu)}{\omega}A(\nu)A(\nu+\omega)$ ($n_F(\nu)=\frac{1}{e^{\beta\nu}+1}$), which strongly validates the Keldysh approach above focusing solely on variations induced by $\varphi$ fields to extract the DC conductivity irregardless of Fermi liquid or incoherent metal regime.

\subsection{Effective action for TRP $\epsilon$ fields and energy correlations}
The expected low energy diffusive modes consist of those induced by phase fluctuations and time reparametrization (TRP) $\tau\rightarrow f(\tau)=\tau+\epsilon(\tau)$, which are primarily associated with charge and energy transport, respectively.   Strictly speaking, while we continue to use the nomenclature, following earlier work on SYK models, of time-reparametrization symmetry, the true symmetry of the problem is time-translation only, which is what generally leads to energy conservation.    

We are guided by the principles of hydrodynamics.  At low frequency and momentum, energy behaves as a diffusive mode, which constrains its response function completely in terms of thermodynamics and kinetic coefficients.  This implies the low energy action {\em must} take the form ($\bar Z=\int [d\epsilon] e^{iNS_\epsilon}$)
\begin{equation}
  \label{eq:5}
iS_\epsilon  =2 \gamma T\int_{-\infty}^{+\infty} d\omega (i\omega^2-p^2 \omega D_{\epsilon} )\epsilon_{c,\omega}\epsilon_{q,-\omega} + \cdots,
\end{equation}
neglecting as usual purely quantum ($\epsilon_q^2$) terms.  Here $\gamma = C/(NT)$ is the Sommerfeld coefficient if temperature is small, and $D_\epsilon$ defines the energy diffusion constant.  Our task is to identify the origin of the two terms in the effecive action above and determine the coefficients.  

First, we note that $\gamma$ is purely thermodynamic in origin, and so can be extracted directly from the free energy.  This is simple and direct, and preferable to a subtle extraction of the specific heat from fluctuations of the energy.  It remains to determine the second term, whose coefficient gives the product $\gamma D_\epsilon$.  To see how this arises, we consider the variation of composite fields under TRP:
 \begin{eqnarray}
 \label{rep_variation1}
t_s\rightarrow  f_s(t_s)=t_s+\epsilon_s(t)\quad G_{ss',x}^f(t_1,t_2)\rightarrow G_{ss',x}[f_s(t_1),f_{s'}(t_2)],
 \end{eqnarray}
which for small $\epsilon$ gives 
\begin{eqnarray}
 \label{rep_variation}
\delta_\epsilon G_{ss',x}(t_1,t_2)=(\epsilon_{s,x}(t_1)-\epsilon_{s',x}(t_2))\partial_tG_{ss',x}(t_1,t_2)+\frac{1}{2}(\epsilon_{s,x}(t_1)-\epsilon_{s',x}(t_2))^2 \partial_t^2 G_{ss',x}(t_1,t_2),
 \end{eqnarray}
where $t=t_1-t_2$.  The idea is, as we did for charge, to include energy fluctuations $\epsilon_s(t)$ through the replacement in \eqref{rep_variation1}, maintaining $G_{ss',x}(t_1,t_2)$ on the right-hand side in the saddle point form, $G_{ss'}(t_1-t_2)$.  The expression for the variation of $G$ in \eqref{rep_variation} was simplified using this assumption.  In this way we are able to identify specifically the modes associated to energy, without other superfluous degrees of freedom.  Please note that the transformation given in \eqref{rep_variation1} differs from the conformal one \eqref{time_rep} in that we do not include the prefactor involving $f'_s(t)$, replacing this by unity.  Due to the lack of conformal symmetry (due to its breaking by the hopping), the scaling dimension of the fermion is ambiguous.  So we intentionally neglect these factors, which calculation shows results in missing some terms of $O[(\epsilon'_s)^2]$ and higher.  Comparing to \eqref{eq:5}, we see that this means we cannot extract in this way the $\gamma$ term, but are still able to obtain the desired $\gamma D_\epsilon$ one.  This is acceptable since we obtain $\gamma$ from an independent thermodynamic calculation.

By inserting the above transformation in the Keldysh action, we may derive, subject to the caveat just outlined, the effective action for TRP modes.   Since we seek only the momentum-dependent part of $S_\epsilon$, it is sufficient to consider only the variation of the hopping term.  This is because all other terms in \eqref{keldysh_action} are fully local in $x$, and so cannot contribute any gradients: they are fully invariant under {\em independent} local (but constant in time) time translations at every $x$.  

The variation of the hopping term is, to quadratic order, (omitting products of $\epsilon_q\epsilon_q$),
 \begin{eqnarray}
 \label{rep_action_hopping}
 &&iS_{t_0,\epsilon}=-\sum_{\bra xx'\ket,ss'}\int_{-\infty}^{+\infty} dt_1dt_2 t_0^2 ss' [G_{ss',x}(t)+\delta_\epsilon G_{ss',x}(t)][G_{s's,x'}(-t)+\delta_\epsilon G_{s's,x'}(-t)]=\text{const}-\sum_{\bra xx'\ket,ss'}\frac{1}{2}ss't_0^2\partial_t G_{ss',x}(t)\partial_{-t}G_{s's,x'}(-t)\nonumber\\
 &&[\epsilon_{s,x}(t_1)-\epsilon_{s',x}(t_2)+\epsilon_{s',x'}(t_2)-\epsilon_{s,x'}(t_1)]^2=\text{const}+\frac{1}{2}\sum_{{\bf p}} \upsilon({\bf p})t_0^2\int _{-\infty}^{+\infty}d\omega \epsilon_{c,\omega} [4\pkpr(-\omega)-4\pkpr(0)]\epsilon_{q,-\omega}
 \end{eqnarray}
 where we have written $\partial_t^2G_{ss',x}(t) G_{s's,x+1}(-t) [\epsilon_{s,x}(t_1)-\epsilon_{s',x}(t_2)]$ as $\partial_tG_{ss',x}(t)\partial_{-t} G_{s's,x+1}(-t) [\epsilon_{s,x}(t_1)-\epsilon_{s',x}(t_2)]$ by integration by parts and ignoring derivatives of $\epsilon$ since they are of higher order (they can contribute to the $\gamma$ term but we get that independently), and switched to Fourier space (taken continuum limit), used $\epsilon_{c/q}$ variables in the last identity. 
 
 We have
$\tilde\pkpr(t)=2i\partial_tG_K(t)\text{Im}[\partial_{-t}G_R(-t)] $
and $P_{3}(\omega)$ is Fourier transform of $\tilde P_{3}(t)$. This is the second part in \eqref{eq:5} and combining with the first part we get \eqref{ep_action}.
 
 We verify that the kernel in \eqref{ep_action} $\Lambda_3(\omega)=\frac{-t_0^2}{2}[4\pkpr(-\omega)-4\pkpr(0)] $ in   vanishes exactly at zero-frequency,  approximate $\Lambda_3(\omega)\approx 2\gamma T^2 D_\epsilon\omega$ ($D_\epsilon$ is the energy diffusion constant), take $\upsilon({\bf p})={\bf p}^2$  and get the effective action for TRP modes
 \begin{equation}
 iS_{\epsilon}=\sum_p\int_{-\infty}^{+\infty}d\omega ( 2i\gamma\omega^2T^2-p^2\Lambda_3(\omega))\epsilon_{c,\omega}\epsilon_{q,-\omega}=2\gamma T^2\int_{-\infty}^{+\infty} d\omega (i\omega^2-p^2 \omega D_{\epsilon} )\epsilon_{c,\omega}\epsilon_{q,-\omega}.
 \end{equation}
 
We further obtain $\varepsilon_{c/q}=\frac{iN\delta S_\epsilon}{\delta \dot\epsilon_{c/q}}=(\varepsilon_+\pm\varepsilon_-)\approx2iN\gamma T^2\dot \epsilon_{q/c}$. The correlator is related as (supplementary information)
\begin{equation}
D_{R\varepsilon}(p,\omega)=\frac{i}{2}\bra \varepsilon_c(x,t)\varepsilon_q(0,0)\ket=-i2N^2\gamma^2 T^4\omega^2 \bra \epsilon_q\epsilon_c\ket_{p,\omega}+N\gamma T^2=\frac{-iN\gamma T^2\omega}{i\omega-D_\epsilon p^2}+N\gamma T^2=\frac{-N\gamma T^2D_{\epsilon}p^2}{i\omega-D_{\epsilon}p^2}.
\end{equation}

One could extract $2N\gamma D_\epsilon=2\kappa/T$, by calculating $\lim_ {\omega\rightarrow 0} \frac{\Lambda_3(\omega)}{\omega}$.

  
 \section{Numerical approach to Calculating real-time Green's functions at $\mu=0$}
 \label{numerics_green}
We aim to solve for the real-time Green's functions iteratively starting from the Green's function in the conformal limit.
The self-energies $\Sigma$   satisfy after the Keldysh rotation the matrix equation $G(\omega)(\omega-\Sigma(\omega))=1$, with
\begin{eqnarray}
\label{sigma_keldysh}
&&\Sigma=L\Sigma \sigma^z L^\dagger\quad \Sigma=\left (\begin{array}{cc} \Sigma_R & \Sigma _K\\ 0 &\Sigma_A\end{array}\right)=\frac{1}{2}\left (\begin{array}{cc} \Sigma_{+-}+\Sigma_{++}-\Sigma_{--}-\Sigma_{-+} & -\Sigma_{+-}+\Sigma_{++}+\Sigma_{--}-\Sigma_{-+} \\ 0 &-\Sigma_{+-}+\Sigma_{++}-\Sigma_{--}+\Sigma_{-+}\end{array}\right).
\end{eqnarray}


In the conformal limit, $\Sigma_R(t)=\frac{i\sqrt 2 b^3}{\sinh(\pi t/\beta)^{3/2}}\theta(t),
\Sigma_K(t)=\frac{\text{sgn}(t)\sqrt 2 b^3}{\sinh(\pi t/\beta)^{3/2}}$ 
where $b=\frac{\pi^{1/4}}{\sqrt {2\beta U_0}}$.  However, the na\"ive imaginary part of $\Sigma_R(\omega)$ (the Fourier transform of the conformal limit self-energy $\Sigma_R(t)$) is divergent for small frequency and which inhibits convergence in the brute force iteration process, i.e., if we na\"ively use the matrix equation $G(\omega)[\omega-\Sigma(\omega)]=1$ to update $G_R(\omega)$ from $\Sigma_R(\omega)$.  A second issue is with a na\"ive iteration is that the self-consistent equations do not specify the temperature, and the final converged solution might correspond to a different temperature than the initial one.

To circumvent these problems, we use the fluctuation-dissipation relation (see Supplementary Information) 
\begin{equation}
\label{fluctuation-dissipationT}
G_K(\omega)=2i\tanh(\frac{\beta\omega}{2})\text{Im}[G_R(\omega)]
\end{equation}
to get a closed set of equations of the Keldysh and retarded components.  Using this relation guarantees we obtain a true equilibrium solution at temperature $1/\beta$.
 
We select the following two equations involving $G_R,G_K,\Sigma_R,\Sigma_K$ to update on $G_R,G_K$ during iteration.
\begin{eqnarray}
\label{eqset}
\frac{1}{\omega-\text{Re}\Sigma_R(\omega)-i\text{Im}\Sigma_R(\omega)}=\text{Re}[G_R(\omega)]+i \text{Im}[G_R(\omega)],\quad G_K(\omega)=[\text{Re}[G_R(\omega)]^2+\text{Im}[G_R(\omega)]^2]\Sigma_K(\omega)
\end{eqnarray}
where the second equation comes from the matrix equation 
\begin{eqnarray}
&&-G_R(\omega)\Sigma_K(\omega)+G_K(\omega)(\omega-\Sigma_A(\omega))=0,\quad G_A(\omega)(\omega-\Sigma_A(\omega))=1.\nonumber
\end{eqnarray} 

From the above equations \eqref{fluctuation-dissipationT}\eqref{eqset}, after some algebra, one finds 
\[ \text{Re}[G_R(\omega)]=\frac{4\tanh(\frac{\beta\omega}{2})^2 \Upsilon(\omega)}{\Sigma_K(\omega)^2-\tanh(\frac{\beta\omega}{2})^2\Upsilon(\omega)^2},\quad \text{Im}[G_R(\omega)]=\frac{i\Sigma_K(\omega) \text{Re}[G_R(\omega)]}{2\Upsilon(\omega)\tanh(\beta\omega/2)},
\]
where $\Upsilon(\omega)\equiv \text{Re}[\Sigma_R(\omega)]-\omega$. From this, we  avoid the divergence in $\text{Im}\Sigma_R$ at the conformal limit and can successfully perform the iteration. (For more details see Supplementary Information.)



%
%
 \section{Thermodynamics: entropy and compressibility}
 \label{thermodynamics}
  
    \begin{figure}[htbp] 
 \begin{center}
  \includegraphics[width=0.45\textwidth]{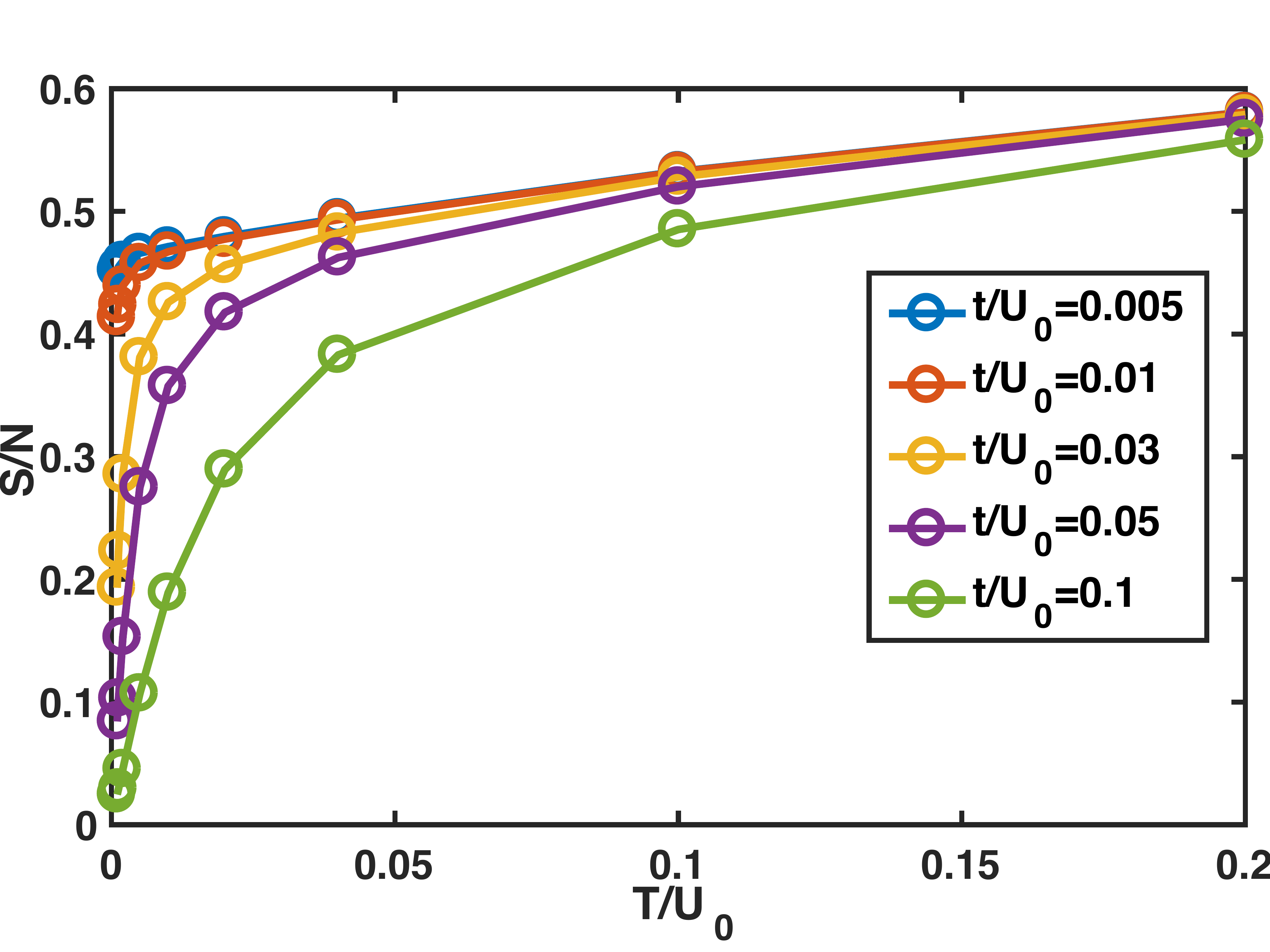}
   \includegraphics[width=0.43\textwidth]{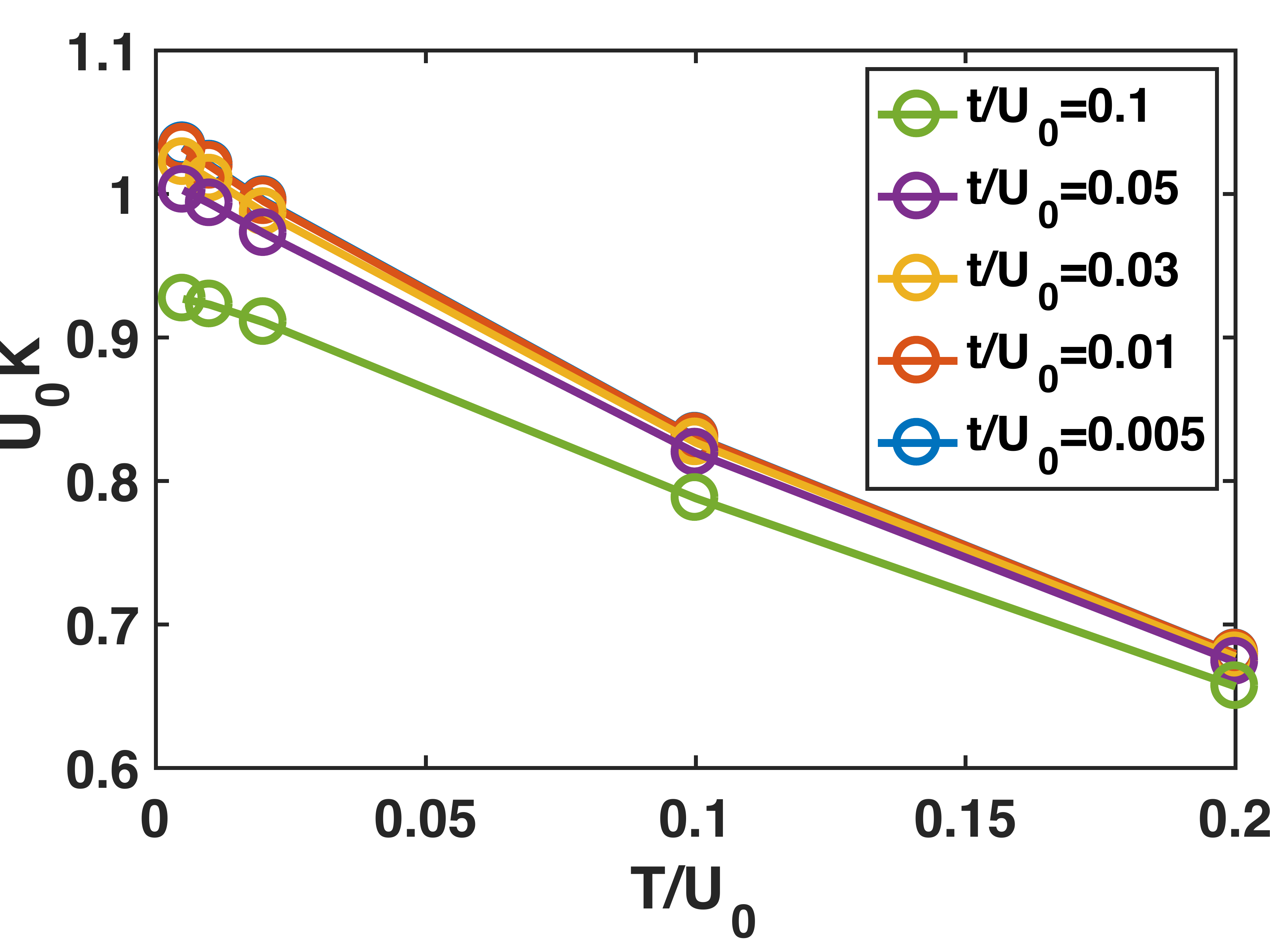}
 \caption{$t$ in the legend denotes $t=\sqrt{\frac{z}{2}} t_0$ where $z$ is the coordination number.  Left: The entropy for $\text{SYK}_4$ (i.e. vanishing $t_0$) agrees with the results in Ref~\onlinecite{fu_2016} and entropy approaches identically regardless of $t/U_0$ the universal $\ln 2$ for high temperature (not shown in the figure). For nonzero $t$, entropy tends to vanish identically as temperature $T\rightarrow 0$. Right: The compressibility $K$ extracted by virtue of $K=\frac{1}{N}\frac{\partial \mathcal N}{\partial \mu}$. For $t/U_0\ll 1$ and relatively small temperature (i.e., $T\ll U_0$), $K$ is  independent of temperature and varies little with $t$. }    \label{thermo_re}
 \end{center}
 \end{figure}
  In the large N limit, the grand canonical potential $\Omega$ is approximated by the saddle point action value of \eqref{euc_action_fermion}, and by virtue of equation of motion \eqref{SD}, can be written as 
  \begin{eqnarray}
 \label{FE}
&& \frac{\Omega}{N}=\frac{-\ln Z}{N\beta}=\frac{S}{\beta N}=T\Big [\sum_n \ln[G(i\omega_n)/G_0(i\omega_n)]-\frac{3}{4}\sum_n \Sigma(i\omega_n)G(i\omega_n)+\frac{zt_0^2}{4}\sum_n G(i\omega_n)^2-\ln(1+e^{\mu/T})\Big ],
 \end{eqnarray}
 where $z$ is the coordination number of the lattice under consideration and we have regularized the free energy by subtracting the part for free fermion, i.e.,$G_{0}(i\omega_n)=\frac{1}{i\omega_n}$, and adding back $-T\ln (1+e^{\mu/T})$. One switches to Helmholtz free energy which depends on ``universal" particle number density $\mathcal N/N$ by a legendre transformation $\mathcal F=\Omega/N +\mu \mathcal N/N$, and obtain entropy density by $\mathtt S/N=\frac{-\partial\mathcal F}{\partial T}$. The entropy for $\text{SYK}_4$ (i.e. vanishing $t_0$) agrees with the results in Ref~\onlinecite {fu_2016} and entropy (Fig \ref{thermo_re}) approaches identically regardless of $t_0/U_0$ the universal $\ln 2$ for high temperature (not shown in the figure). The entropy is significantly reduced for small temperature by the presence of two-fermion hopping. 
 
 The compressibility is obtained as  $K=\frac{1}{N}\frac{\partial \mathcal N}{\partial \mu}$ or $K=-1/(\frac{\partial^2\mathcal F}{\partial^2 \frac{\mathcal N}{N}})$. The plot in Fig.~\ref{thermo_re} shows the results using the first derivative method (which agrees with that found in Ref~\onlinecite{davidson_2016} as well as a large-$q$ calculation (unpublished)).

%

 ~
 ~
 ~
 ~
 \newpage
  \appendix
{\Large Supplementary Information for \textit{A strongly correlated metal built from Sachdev-Ye-Kitaev models}}

\section{Details of Keldysh}
\label{sec:details-keldysh}

\subsection{Fermion Green's functions}
\label{sec:ferm-greens-funct}

Explicitly writing down the expression for $G$, using the Keldysh
ordering of times on the contour, one finds they evaluate to following objects:
\begin{eqnarray}
\label{g_object}
&&G_{++}(t,t')=-i\bra T(c(t)c^\dagger(t'))\ket,\qquad G_{+-}(t,t')=i\bra c^\dagger(t')c(t)\ket,\nonumber\\
&&G_{-+}(t,t')=-i\bra c(t) c^\dagger(t')\ket,\qquad G_{--}(t,t')=-i\bra T^{-1}(c(t) c^\dagger(t'))\ket.
\end{eqnarray}
From the above, it is clear that the four functions are not
independent, and indeed they are constrained by the identity,
following from Eq.~\eqref{g_object}, that $ G_{++}+ G_{--}- G_{+-}- G_{-+}=0$.

Analytic continuation of the imaginary time Green's function is possible when we know the latter fully.  We have
\begin{eqnarray}
 G_{-+}(t)=\begin{cases} i G(\tau>0\rightarrow it), & t>0 \\
 i G(\tau>0 \rightarrow it), &t<0\end{cases}, \qquad
 G_{++}(t) =\begin{cases} i G(\tau>0\rightarrow it) & t>0\\
i G(\tau<0\rightarrow it) & t<0\end{cases}\nonumber\\
 G_{+-}(t)=\begin{cases} i G(\tau<0\rightarrow it), & t>0 \\
 i G(\tau<0 \rightarrow it), &t<0\end{cases}, \qquad
  G_{--}(t) =\begin{cases} i G(\tau<0\rightarrow it) & t>0\\
i G(\tau>0\rightarrow it) & t<0\end{cases}.
\end{eqnarray}
This gives
\begin{eqnarray}
\label{G_re}
&&G_{++}(t)=\begin{cases} -ib\frac{e^{-i\frac{\pi}{4}}}{\sqrt{\sinh[\frac{\pi t}{\beta}]}}& (t>0)\\
ib\frac{e^{-i\frac{\pi}{4}}}{\sqrt{\sinh[\frac{\pi |t|}{\beta}]}}& (t<0)\end{cases}\quad
G_{-+}(t)=\begin{cases} -ib\frac{e^{-i\frac{\pi}{4}}}{\sqrt{\sinh[\frac{\pi t}{\beta}]}}& (t>0)\\
-ib\frac{e^{i\frac{\pi}{4}}}{\sqrt{\sinh[\frac{\pi |t|}{\beta}]}}& (t<0)\end{cases}\nonumber\\
&&G_{--}(t)=\begin{cases} ib\frac{e^{i\frac{\pi}{4}}}{\sqrt{\sinh[\frac{\pi t}{\beta}]}}& (t>0)\\
-ib\frac{e^{i\frac{\pi}{4}}}{\sqrt{\sinh[\frac{\pi |t|}{\beta}]}}& (t<0)\end{cases}\quad
G_{+-}(t)=\begin{cases} ib\frac{e^{i\frac{\pi}{4}}}{\sqrt{\sinh[\frac{\pi t}{\beta}]}}& (t>0)\\
ib\frac{e^{-i\frac{\pi}{4}}}{\sqrt{\sinh[\frac{\pi |t|}{\beta}]}}& (t<0)\end{cases}.
\end{eqnarray}
Combining the above results using the definition of the Keldysh rotation gives the results in the Methods section.

\subsection{Density correlations}
\label{sec:density-correlations}

In the main text, we asserted that the the retarded density correlator
only depends on the terms of the form $\varphi_c\varphi_q$ in the
effective action.  This is because the above full action is of the form
\[ S=\int_0^{+\infty} d\omega (\varphi_{c,\omega},\varphi_{q,-\omega})\left(\begin{array}{cc} 0& P(\omega) \\ P(-\omega)& R(\omega)\end{array}\right) \left(\begin{array}{c}\varphi_{c,-\omega}\\ \varphi_{q,\omega}\end{array}\right).\]
Notably, because a pure classical variation of the phase field
$\varphi_c$ does not change the partition function, hence products of
$\varphi_c,\varphi_c$ vanish. The inverse of a  matrix of the above
form is
\[\left(\begin{array}{cc} 0& a\\b &c\end{array}\right)^{-1}=\left(\begin{array}{cc}\frac{-c}{ab}& \frac{1}{b}\\ \frac{1}{a} & 0\end{array}\right) .\] 
Note that the off-diagonal elements of the inverse only depend on
$a,b$. As we shall see below the retarded density-density correlator
depends on $\bra \varphi_{c,\omega}\varphi_{q,-\omega}\ket,\bra
\varphi_{q,\omega}\varphi_{q,-\omega}\ket$.  The form of the inverse
implies that the expectation value $\bra
\varphi_{q,\omega}\varphi_{q,-\omega}\ket$ vanishes. Hence products of
$\varphi_q$ in the action for $\varphi$ are not of interest here, as
they do not contribute to $\bra \varphi_{c,\omega}\varphi_{q,-\omega}\ket$.

To obtain the retarded density-density correlator, we obtain first the relations
\begin{eqnarray}
\bra  \mathcal N_+(x,t)\mathcal N_+(x',t')\ket =\bra \mathcal T\mathcal N(x,t)\mathcal N(x',t')\ket\quad
\bra\mathcal N_+(x,t)\mathcal N_-(x',t')\ket=\bra\mathcal N(x',t')\mathcal N(x,t)\ket\nonumber\\
\bra  \mathcal N_-(x,t)\mathcal N_-(x',t')\ket =\bra \mathcal T^{-1}\mathcal N(x,t)\mathcal N(x',t')\ket\quad
\bra\mathcal N_-(x,t)\mathcal N_+(x',t')\ket=\bra\mathcal N(x,t)\mathcal N(x',t')\ket.
\end{eqnarray}
From this, we verify the following identity
\begin{equation}
D_{Rn}(x,t)\equiv i\theta(t)\bra[\mathcal N(x,t),\mathcal N(0,0)]\ket =\frac{i}{2}\bra \mathcal N_+\mathcal N_+-\mathcal N_+\mathcal N_-+\mathcal N_-\mathcal N_+-\mathcal N_-\mathcal N_-\ket=\frac{i}{2}\bra \mathcal N_c(x,t)\mathcal N_q(0,0)\ket,
\end{equation}
where $\mathcal N_s\equiv \frac{N\delta S_\varphi}{\delta \dot\varphi_s}, \mathcal N_{c/q}=\frac{N\delta S_\varphi}{\delta \dot\varphi_{c/q}}=(\mathcal N_+\pm\mathcal N_-)$. From the expression of $\mathcal N_{c/q}$ in the Methods, we verify that the retarded correlator only depends on $\bra \varphi_{c,\omega}\varphi_{q,-\omega}\ket,\bra \varphi_{q,\omega}\varphi_{q,-\omega}\ket$.  

Similarly for energy density correlators, using the relations
\begin{eqnarray}
\bra  \varepsilon_+(x,t)\varepsilon_+(x',t')\ket =\bra \mathcal T\varepsilon(x,t)\varepsilon(x',t')\ket,\quad
\bra\varepsilon_+(x,t)\varepsilon_-(x',t')\ket=\bra\varepsilon(x',t')\varepsilon(x,t)\ket,\nonumber\\
\bra  \varepsilon_-(x,t)\varepsilon_-(x',t')\ket =\bra \mathcal T^{-1}\varepsilon(x,t)\varepsilon(x',t')\ket,\quad
\bra\varepsilon_-(x,t)\varepsilon_+(x',t')\ket=\bra\varepsilon(x,t)\varepsilon(x',t')\ket,
\end{eqnarray}
we verify the following identity
\begin{equation}
D_{R\varepsilon}(x,t)\equiv i\theta(t)\bra[\varepsilon(x,t),\varepsilon(0,0)]\ket =\frac{i}{2}\bra \varepsilon_+\varepsilon_+-\varepsilon_+\varepsilon_-+\varepsilon_-\varepsilon_+-\varepsilon_-\varepsilon_-\ket=\frac{i}{2}\bra \varepsilon_c(x,t)\varepsilon_q(0,0)\ket.
\end{equation}
Here $\varepsilon_s\equiv \frac{iN\delta S_{\epsilon}}{\delta \dot\epsilon_s}, \varepsilon_{c/q}=\frac{iN\delta S_\epsilon}{\delta \dot\epsilon_{c/q}}=(\varepsilon_+\pm\varepsilon_-)$.

\subsection{Fluctuation-dissipation relation}
The set of Green's functions are all be determined by the spectral
function for the equilibrium system at a fixed temperature. Define following spectral functions:
\begin{eqnarray}
A^+(\omega)=e^{\frac{\beta\omega}{2}}\sum_{n,m}\bra n|C_a|m\ket\bra m|C_a^\dagger|n\ket \frac{e^{\frac{-(\epsilon_m+\epsilon_n)\beta}{2}}}{Z}\delta[\omega-(\epsilon_m-\epsilon_n)],\nonumber\\
A^-(\omega)=e^{\frac{-\beta\omega}{2}}\sum_{n,m}\bra n|C_a|m\ket\bra m|C_a^\dagger|n\ket \frac{e^{\frac{-(\epsilon_m+\epsilon_n)\beta}{2}}}{Z}\delta[\omega-(\epsilon_m-\epsilon_n)],
\end{eqnarray}
where $\epsilon_m\equiv \varepsilon_m-\mu \mathcal N_m$, with $m,n$
labels for the many-body eigenstates of the system, and $\varepsilon_m, \mathcal N_m$ denote the energy/particle number of the eigenstate.

The retarded and advanced Green's function are expressed as
\begin{eqnarray}
&&G_{R/A}(\omega)=\int_{-\infty}^{+\infty} d\nu \frac{A^+(\nu)+A^-(\nu)}{\omega-\nu\pm i0^+},\qquad G_R(\omega)^*=G_A(\omega).
\end{eqnarray}
Similarly,  the spectral representation for the Keldysh function
reveals the fluctuation-dissipation relation:
\begin{eqnarray}
\label{krrelation}
&&iG_K(t,t')=\sum_{n,m}\bra n|C_a|m\ket\bra m|C_a^\dagger|n\ket e^{-i(\epsilon_m-\epsilon_n)(t-t')}(e^{-\beta\epsilon_n}-e^{-\beta\epsilon_m}),\qquad G_K(\omega)=2i \tanh(\frac{\beta\omega}{2}) \text{Im} G_R(\omega).
\end{eqnarray}

\section{Computing $\sigma/\kappa$ at SYK$_4$/free fermion limit and comparison with numerical results}
 \label{analytical}
 \subsection{Electric conductivity at SYK$_4$ and free fermion limit}
We analytically calculate the kernel $\Lambda_2(\omega)$ in the SYK$_4$ limit by using the conformal results in \eqref{G_re}:
\begin{eqnarray}
\Lambda_2(\omega)_{\text{SYK}_4}=-4t_0^2\int_0^{\infty} dt \text{Im}[G_K(-t)G_R(t)]\sin(\omega t)=8t_0^2b^2\int_0^\infty dt \frac{\sin[\omega t]}{\sinh[\pi t/\beta]}=4t_0^2b^2\beta\tanh[\frac{\beta\omega}{2}],
\end{eqnarray}
with $b=\frac{\pi^{\frac{1}{4}}}{\sqrt{2\beta U_0}}$ as given before
in the Methods section.  
As shown in the main text, we obtain $2KD_{\varphi} =
\lim_{\omega\rightarrow 0} \frac{\Lambda_2(\omega)}{\omega}$, which
gives
\begin{equation}
  \label{eq:3}
  \frac{\sigma_{\text{SYK}_4}}{N}=\frac{\sqrt {\pi}\beta t_0^2}{2 U_0}.
\end{equation}
This evaluates to $\sigma_{\text{SYK}_4}/N \approx 0.886\frac{E_c}{T}$, while in numerics we have $\sigma/N\approx 0.88 \frac{E_c}{T}$.

One can obtain the same results in the frequency domain.  From the expression for $\Lambda_2(\omega)$, we have (using $G_K(\omega)=2i\tanh(\beta\omega/2)\text{Im}G_R(\omega)$)
\begin{eqnarray}
\label{s22}
2\sigma/N&=&\lim_{\omega\rightarrow 0}\frac{\Lambda_2(\omega)}{\omega}=\frac{2t_0^2}{2\pi}\frac{d}{d\omega}\int_{-\infty}^{+\infty} d\omega_1 G_K(\omega_1)G_R(\omega_1+\omega)-G_K(\omega_1)^*G_R(\omega_1-\omega)^*=\frac{-2t_0^2}{\pi}\int_{-\infty}^{+\infty} d\omega \tanh(\beta\omega/2)\frac{\partial \text{Im}G_R^2}{\partial\omega}\nonumber\\
&=&\frac{t_0^2}{\pi}\int_{-\infty}^{+\infty} dx \text{sech}(x/2)^2 (\text{Im}G_R(x/\beta))^2,
\end{eqnarray}
where we have integrated by parts in the second line ($x=\beta \omega$).

Inserting the conformal limit expression for ${\rm Im} G_R(\omega)$
from the Methods section,
we then find from the above equation
\begin{equation}
2\sigma_{\text{SYK}_4}/N=\frac{\beta t_0^2}{2\pi \sqrt \pi U_0}\int_{-\infty}^{+\infty} dx \text{sech}(x/2)^2  \text{Re} [\frac{\Gamma(\frac{1}{4}-\frac{ix}{2\pi})}{\Gamma(\frac{3}{4}-\frac{ix}{2\pi})}]^2\approx \frac{1.772E_c}{T}.
\end{equation}

Now we turn to the fermion limit.  There the analytical solution to
the Green's function is obtained (we define $\tilde t_0\equiv
\frac{\sqrt z}{\sqrt 2} t_0$ throughout the supplementary information) as
 \begin{eqnarray}
 &&G_R(\omega)=\frac{-\omega-\mu+\text{sgn}(\omega)\sqrt{(\omega+\mu)^2-8\tilde t_0^2}}{-4\tilde t_0^2}.
 \end{eqnarray}
We use \eqref{s22} to arrive at
 \begin{eqnarray}
 \label{sigma_cf}
&&  \frac{\sigma}{N}
=\int_{-2\sqrt 2 \tilde t_0}^{2\sqrt 2 \tilde t_0}
   d\omega_1\frac{\beta}{32\pi \tilde t_0^2}(8\tilde
   t_0^2-\omega_1^2){\rm sech}^2\left(\frac{\beta \omega_1}{2}\right).
 \end{eqnarray}
At zero temperature, this simplifies to ($\sigma_{sat}$ denotes the saturated value of $\sigma$ in the free fermion limit)
 \begin{eqnarray}
 \label{0T}
 \frac{\sigma_{sat}}{N}=\frac{1}{\pi},
 \end{eqnarray}
 which agrees with the numerics ($\sigma_{sat}/N\approx 0.32$ in numerics) for the case $t_0/U_0\ll 1, T\ll E_c$.

\subsection{Thermal conductivity at SYK$_4$ and free fermion limit}
  We calculate $\kappa/T$ in the SYK$_4$ conformal/free fermion
  limit. Recall $\kappa = \frac{N}{2T}\lim_{\omega\rightarrow 0}
  \frac{\Lambda_3(\omega)}{\omega}$.  We obtain this from
  $\Lambda_3(\omega)=\frac{-t_0^2}{2}[4\pkpr(-\omega)-4\pkpr(0)]$,
  where 
  \begin{equation}
    \label{eq:1}
    P_3(\omega)= \frac{1}{2\pi}\int_{-\infty}^{+\infty} d\omega_1[-G_R(\omega_1-\omega)G_K(\omega_1) \omega_1(\omega_1-\omega)-G_R^*(\omega_1+\omega)G_K(\omega_1)\omega_1(\omega_1+\omega)].
  \end{equation}
Taking the limit, we have
 \begin{eqnarray}
 \label{kappaex}
\frac{\kappa}{NT}&=&\frac{-t_0^2}{T^2}\frac{\partial P_3(-\omega)}{\partial \omega}\Big |_{\omega=0}=\frac{i\beta^2 t_0^2}{2\pi}\int _{-\infty}^{+\infty}d\omega_1 [2G_K\text{Im}G_R \omega_1+2G_K\frac{\partial\text{Im} G_R}{\partial \omega} \omega_1^2]\nonumber\\
&=&\frac{-4\beta^2 t_0^2}{2\pi}\int_{-\infty}^{+\infty} d[\frac{\text{Im}[G_R]^2\omega^2}{2}] \tanh(\frac{\beta\omega}{2})=\frac{t_0^2}{2\pi}\int_{-\infty}^{+\infty} \! dx\, \text{sech} (x/2)^2 (\text{Im}G_R(\frac{x}{\beta})^2 x^2)
 \end{eqnarray}
where we have used $\lim_{\omega\rightarrow \infty} (\omega\text{Im}G_R(\omega)) = 0$ and integrated by parts ($x=\beta\omega$).

 One can directly evaluate this numerically using the conformal limit form of $G_R$:
 \begin{eqnarray}
 \label{kappa_cf}
 \frac{\kappa_{\text{SYK}_4}}{NT}
 =\frac{-\beta t_0^2}{4\sqrt \pi U_0}\int_{-\infty}^{+\infty}  dx \text{sech}(x/2)^2 \frac{x^2}{\pi}\text{Re}[\frac{\Gamma(\frac{1}{4}-\frac{ix}{2\pi})}{\Gamma(\frac{3}{4}-\frac{ix}{2\pi})}]^2\approx 1.092\frac{E_c}{T}
 \end{eqnarray}
where we have plugged in $G_R(\omega)=\frac{-i\sqrt{\beta}}{\sqrt{2U_0\sqrt{\pi}}}\frac{\Gamma(\frac{1}{4}-\frac{i\beta\omega}{2\pi})}{\Gamma(\frac{3}{4}-\frac{i\beta\omega}{2\pi})}\rightarrow\frac{-i \pi^{\frac{1}{4}} e^{i\frac{\pi}{4}}}{\sqrt {U_0\omega}}$ when $\beta\omega$ is relatively large, in the conformal limit. In numerics we have in $E_c\ll T\ll U_0$, $\frac{\kappa}{NT}\approx 1.08\frac{E_c}{T}$. 

We obtain an analytic result by working instead in real time. From $P_3(t)=2i\partial_tG_K(t)\text{Im}[\partial_{-t}G_R(-t)]$, one has $\frac{\partial P_3(\omega)}{\partial \omega}|_{\omega=0}\sim \int dt \partial_t G_K(t)\partial _{-t} \text{Im}G_R(-t) t$, plugging in the conformal limit Green's function \eqref{G_re}, the integral $\sim \int dt \frac{t}{\sinh(\pi t/\beta)^3}$ diverges due to the singular behavior of Green's function at small time (since conformal limit is only valid for small frequency, i.e., long time scale). To regularize the divergence, we add to the integral a total derivative and write down
\begin{eqnarray}
\frac{\kappa_{\text{SYK}_4}}{NT}=2\beta^2t_0^2\int dt  \partial_t G_K(t)\partial _{-t} \text{Im}G_R(-t) t-\beta^2 t_0^2 \int dt \partial_t[G_K(-t)\text{Im}G_R(t)]\nonumber\\
=b^2t_0^2\beta^2\int_0^{+\infty} d(\frac{\pi t}{\beta}) \frac{\text{cosh}(\frac{\pi t}{\beta})(\frac{\pi t}{\beta}\text{coth} (\frac{\pi t}{\beta})-1)}{\text{sinh}(\frac{\pi t}{\beta})^2}=\frac{\pi^{\frac{5}{2}}E_c}{16 T}.
\end{eqnarray}
where we used the fact that $\int_0^{+\infty}
dt \partial_t[G_K(-t)\text{Im}G_R(t)]=0$ since the exact (not
conformal limit) UV regularized function $G_K(0)=0, G_R(0)$ is
finite.  After the manipulations, the integral in the second line is
convergent in the SYK$_4$ limit and can be done analytically.  This
gives the exact result in the incoherent regime.  Quantitatively this
agrees within numerical accuracy with the result of the frequency
integral above.

Returning to the free fermion limit, we substituting the SYK$_2$ limit of Green's function $\text{Im}G_R(\omega)=\theta(2\sqrt 2 \tilde t_0-|\omega|)\frac{-\sqrt{8\tilde t_0^2-\omega^2}}{4\tilde t_0^2}$ in \eqref{kappaex}, one has
 \begin{eqnarray}
 &&\frac{\kappa_{sat}}{NT}=\frac{-4\beta^2\tilde
    t_0^2}{2\pi}\int_{-2\sqrt 2 \tilde t_0}^{2\sqrt 2 \tilde t_0}
    \tanh(\beta\omega/2)[\frac{8\tilde t_0^2-2\omega^2}{16\tilde
    t_0^4}\omega] d\omega. 
\end{eqnarray}
Taking the low temperature limit, we get
\begin{equation}
  \label{eq:2}
  \frac{\kappa_{sat}}{NT}=\frac{\pi}{3}.
\end{equation}


\subsection{Lorentz ratio and diffusion constants}
The Lorentz ratio in the incoherent metal regime using conformal limit result reads
\begin{equation}
\frac{\kappa_{\text{SYK}_4}}{T\sigma_{\text{SYK}_4}}=\frac{\pi^2}{8}\approx1.23 ,
\end{equation}
while in numerics we get $1.2$ for this value. In incoherent metal
regime, $\sigma,\kappa/T\sim E_c/T=t_0^2/(U_0T)$. Since both
$K,\gamma$ in this regime go to their values for the  SYK$_4$
model\cite{maldacena_2016,davidson_2016}, they are comparable, i.e.,
$K,\gamma\sim 1/U_0$.  This implies from the Einstein relations$
\frac{\sigma}{N}=KD_\varphi$ and $\kappa/(NT)=\gamma D_\epsilon$, that
the diffusion constants are comparable, $D_\varphi,D_\epsilon\sim
t_0^2/T$.   Quantitatively, $D_\varphi/D_\epsilon\rightarrow 0.75$ in
the incoherent metal regime.

\begin{table}
\begin{center}
\caption {The analytical and numerical results in conformal / free fermion limit. The column labeled ``SYK$4/2$" lists results using numerical Green's function with $t_0(U_0)=0$ for $\sigma_{\text{SYK$_4$(sat)}},\kappa_{\text{SYK$_4$(sat)}}$, respectively.}
\label{limit_re}
\begin{tabular}{c|ccc|c}
\hline
\hline
 &Analytical&SYK$4/2$ & Numeric & $\frac{\kappa}{T\sigma}$\\
 \hline
 $\sigma_{\text{SYK}_4}/N$ &$\frac{\sqrt \pi E_c}{2T}\approx 0.89\frac{E_c}{T}$ & $0.89\frac{E_c}{T}$ & $0.90 \frac{E_c}{T}$ & $r_{\text{cf,ana}}=\frac{\pi^2}{8}$\\
$ \frac{\kappa_{\text{SYK}_4}}{NT}$ &$\frac{\pi^{\frac{5}{2}}E_c}{16 T}\approx 1.09\frac{E_c}{T}$ & $1.08\frac{E_c}{T}$ & $1.08\frac{E_c}{T}$ & $r_{\text{cf,num}}=1.20$\\
\hline
$\sigma_{\text{sat}}/N$ & $\frac{1}{\pi}\approx 0.318 $ & $0.318$ & $0.32$& $r_{\text{sat,ana}}= \frac{\pi^2}{3}$\\
$\frac{\kappa_{\text{sat}}}{NT}$ &$\frac{\pi}{3}\approx 1.047$ & $1.014$ & $1.045$ & $r_{\text{sat,num}}=0.33\pi^2$\\
\hline
\hline

\end{tabular}
\end{center}
\end{table} 
In the Fermi liquid limit we have
\begin{equation}
\frac{\kappa_{sat}}{T\sigma_{sat}}=\frac{\pi^2}{3}
\end{equation}
which is in line with the Wiedermann-Franz law for conventional
metals. In numerics, this value is $0.33\pi^2$. Since
$\sigma/N=KD_\varphi,\kappa/(NT)=\gamma D_\epsilon$ saturate in this
limit, we have $D_\varphi/D_\epsilon\sim \gamma/K\sim U_0/E_c\gg 1$.
So at low temperature there is a large difference in charge and energy
diffusion constants.  We list the results in the two limits in table \ref{limit_re}.

\section{Numerical methods for real-time Green's function}
\subsection{Self-energy}

From the real time self-consistent equations, expressed in the Keldysh
basis, we obtain ($\tilde t_0\equiv \frac{\sqrt z}{\sqrt 2} t_0$)
\begin{eqnarray}
\label{sigmark}
&&\Sigma_R(t)=2 \tilde t_0^2 G_R(t) + \frac{1}{2} U_0^2 G_K(-t) G_K(t) G_R(t) + \frac{1}{4} U_0^2 G_K(t)^2 G_R^*(t)+ 
 \frac{1}{4} U_0^2 G_R(t)^2G_R^*(t),\nonumber\\
 &&\Sigma_K(t)=\begin{cases} 2 \tilde t_0^2G_K(t)+\frac{1}{4}U_0^2 G_K(-t)G_K(t)^2+\frac{1}{4}U_0^2 G_K(-t)G_R(t)^2+\frac{1}{2}U_0^2 G_K(t)|G_R(t)|^2\quad(t>0)\\
 2 \tilde t_0^2G_K(t)+\frac{1}{4}U_0^2 G_K(-t)G_K(t)^2+\frac{1}{4}U_0^2 G_K(-t)G_R(-t)^{*2}+\frac{1}{2}U_0^2 G_K(t)|G_R(-t)|^2 \quad (t<0)\end{cases}.
 \end{eqnarray}
These equations are used to calculate $\Sigma_{R/K}(t)$.

\subsection{Fourier Transform Algorithm}
To numerically calculate Fourier components of Green's function/self-energies, we discretize time and frequency as an array of points in time $T_{\text{arr}}$ and frequency $\Omega_{\text{arr}}$
\begin{eqnarray}
T_{\text{arr}}=\frac{T_0}{N_t}[0,1,\cdots, N_t-1],\quad
\Omega_{\text{arr}}=\frac{2\pi}{T_{\text{eff}}}[0,1,\cdots, N_{\omega}-1].
\end{eqnarray}

The Fourier transform $G(t)=\frac{1}{2\pi}\int d\omega G(\omega)e^{-i\omega t}$ and $G(\omega)=\int_0^{\infty} dt G(t)e^{i\omega t}$ becomes the discretized version (notice the lower limit of integral in the second equation, which is valid for retarded Green's function and could also be used to calculate Keldysh components by virtue of its property discussed below).  In practice we ``overpad" the time point sequence by a factor of $4$ (points in time sequence $N=4N_t$) to get frequency at which one doesn't an integral number of periods in the $[0,T_0]$ integration range. i.e. $G(T)=[G(T_{\text{arr}}), 0\cdots 0]$ with $3N_t$ $0$s following, and hence $T_{\text{eff}}=4T_0$.
\begin{eqnarray}
\label{fft_formula}
&&G(T(j)>0)=\frac{1}{T_{\text{eff}}}\sum_i G(\Omega(i)>0)e^{-i\frac{2\pi}{N_t}(i-1)(j-1)}+\frac{1}{T_{\text{eff}}}\sum_i G(-\Omega(i)<0)e^{i\frac{2\pi}{Nt}(i-1)(j-1)},\nonumber\\
&&G(\Omega(j)>0)=\frac{T_0}{N_t}\sum_i G(T(i)>0)e^{i\frac{2\pi}{N_t}(i-1)(j-1)},\quad G(-\Omega(j)<0)=\frac{T_0}{N_t}\sum_i G(T(i)>0)e^{-i\frac{2\pi}{N_t}(i-1)(j-1)}.
\end{eqnarray}

This could be performed by {\emph {fft, ifft}} functions in MATLAB. We take $T_0\gg \beta$ and $N_\omega\ll N$ in order for $\Omega(N_\omega) \Delta t\ll 1$ to validate the discretization. Moreover, to make the cutoff in frequency satisfy $G(\Omega(N_\omega) )\ll G(\Omega(1))$, recall that at conformal limit $G(\omega)\sim \frac{1}{\sqrt{\beta \omega}}$, we require $\frac{2\pi N_\omega \beta}{T_0}\gg 1$, which means that one has to update $T_0$ proportional to $\beta$ as we vary $\beta$. 

In numerics, we take 
\begin{equation}
T_0=10\beta, N_t=2^{20}, N_\omega=2^{20}, N=4N_t.
\end{equation}

The above derivation only applies to the retarded components with a factor $\theta(t)$ (hence $T(i)>0$) and for $\Sigma_K(\omega)$, we could use the identity $\Sigma_K(t)=-\Sigma_K(-t)^*$ to modify the algorithm.

\section{Heavy Fermi liquid phenomenology}
 \label{Fermi_liquid}
\subsection{Quasi-particle residue and ``Bad" Fermi liquid} 

The saddle point condition for imaginary-time Green's function is (assuming zero chemical potential,$\tilde t_0\equiv \frac{\sqrt z}{\sqrt 2} t_0,\tilde E_c\equiv \frac{\tilde t_0^2}{U_0}$)
 \begin{eqnarray}
&&G(i\omega)^{-1}=i\omega-\Sigma(i\omega),\quad \Sigma(\tau)=-U_0^2G(\tau)^2G(-\tau)+2\tilde t_0^2G(\tau).
 \end{eqnarray}
 Rescaling functions as
 \begin{eqnarray}
 \label{scale}
 &&\bar \omega=\frac{\omega}{\tilde E_c},\quad\bar \tau=\tau \tilde E_c,\quad \bar G(i\bar\omega)=\tilde t_0G(i\omega),\quad \bar \Sigma(i\bar\omega)=\frac{\Sigma(i\omega)}{\tilde t_0}.
 \end{eqnarray}
 The saddle point equation is formatted as
 \begin{eqnarray}
 \label{scaled}
 \bar G(i\bar\omega)^{-1}=\frac{\tilde E_c}{\tilde t_0}(i\bar\omega-\frac{\tilde t_0}{\tilde E_c}\bar \Sigma(i\bar\omega))\approx \bar \Sigma(i\bar\omega),\nonumber\\
 \bar\Sigma(\bar \tau)=-\bar G(\bar\tau)^2\bar G(-\bar\tau)+2\bar G(\bar\tau),
 \end{eqnarray}
 that, given $\frac{\tilde E_c}{\tilde t_0}\ll 1$, is an equation set
 with only \emph{dimensionless} parameters.  As we argued in the text,
 the low energy behavior is in the realm of Fermi liquid theory.  Then
 the spectral weight $\bar A(\bar \omega)$ 
 should contain a
 quasiparticle contribution, which because it contains no parameters,
 must have a  residue of $O(1)$. From the scaling in \eqref{scale},
 it follows that the width of the ``coherence region" attributed to
 quasiparticle formation in $\bar A(\bar\omega)$ is multiplied by
 $\tilde E_c$ in $A(\omega)$ (i.e. in physical units) and the quasiparticle residue of our
 model (i.e., the integral of $A(\omega)$ within the ``coherence
 region") is
 $Z\sim\frac{\tilde E_c}{\tilde t_0}=\frac{\tilde t_0}{U_0}\ll 1$
 which is characteristic of a ``bad" Fermi
 liquid.
 
 \subsection{Grand canonical potential in Fermi liquid theory, compressibility and Sommerfeld coefficient}
In Landau's Fermi liquid theory, the energy is a functional of a series of ``quasi-particle" states labeled by $a,b$, we have
\begin{equation}
E-\mu\mathcal N=\sum_a \varepsilon_a n_a+\frac{1}{2}\sum_{a,b}f_{ab}(n_a-n_a^0)(n_b-n_b^0)-\mu \sum_a n_a=\text{const}+\sum_a (\varepsilon_a-\bar f\sum_b n_b^0-\mu)n_a+\frac{1}{2}\bar f(\sum_a n_a)^2
\end{equation}
where $n_a,n_a^0$ denotes the occupation number of the quasiparticle state and superscript $0$ denotes the occupation number of the ``reference" state one starts with to define $\varepsilon_a, f_{ab}$, and we take it here to be the state with $\mu=0$,i.e., $\bra n_a\ket_{\mu=0}=n_a^0$. In the second identity we use $\bar f$ to replace $f_{ab}$ for simplicity.

Define $E_a=\varepsilon_a-\bar f\sum_b n_b ^0$, we have for the partition function in grand canonical ensemble as (introduce a hubbard-stratonovich variable $\lambda$)
\begin{eqnarray}
\mathcal Z&=& \sum e^{-\beta(E-\mu \mathcal N)}=\sum_{n_a=0,1} \prod_a e^{-\beta(E_a-\mu)n_a}e^{-\frac{\beta\bar f}{2}(\sum_a n_a)^2}\nonumber\\
&=&\sum_{n_a=0,1} \int \sqrt{\frac{\beta}{2\bar f}} d\lambda e^{-i\beta \lambda\sum_a n_a-\frac{\beta}{2\bar f}\lambda^2}\prod_a e^{-\beta(E_a-\mu)n_a}=\int \sqrt{\frac{\beta}{2\bar f}} d\lambda e^{-\frac{\beta}{2\bar f}\lambda^2}\prod_a [1+e^{-\beta(E_a-\mu+i\lambda})].
\end{eqnarray}

The saddle point condition for $\lambda$ reads
\begin{equation}
\label{sdcon}
\frac{i\lambda_s}{\bar f}=\sum_a \frac{1}{1+e^{\beta (E_a-\mu+i\lambda_s)}}
\end{equation}
and the partition function reads
\begin{equation}
\ln \mathcal Z=\frac{-\beta}{2\bar f}\lambda_s^2+\sum_a \ln[1+e^{-\beta (E_a-\mu+i\lambda_s)}].
\end{equation}

The particle number descends from the derivative of grand canonical potential w.r.t. $\mu$,
\begin{equation}
\mathcal N=\frac{T\partial \ln \mathcal Z}{\partial \mu}\Big |_T=\sum_a \frac{1}{1+e^{\beta (E_a-\mu+i\lambda_s)}}
\end{equation}
where we have used the saddle point condition for $\lambda_s$ and we identify from \eqref{sdcon} that $i\lambda_s=\bar f\mathcal N$.

For compressibility one obtains
\begin{eqnarray}
NK&=&\frac{\partial \mathcal N}{\partial \mu}=\sum_a \frac{\beta e^{\beta(E_a-\mu+i\lambda_s)}}{(1+e^{\beta (E_a-\mu+i\lambda_s)})^2}(1-i\frac{\partial \lambda_s}{\partial \mu})=-\int d\varepsilon_a g(\varepsilon_a) \frac{\partial (\frac{1}{1+e^{\beta(E_a-\mu+i\lambda_s)}})}{\partial \varepsilon_a}(1-\bar f\frac{\partial \mathcal N}{\partial \mu})
\end{eqnarray}
where we write the discrete sum of quasiparticle states as an integral $\int g(\varepsilon_a)d\varepsilon_a$ with $g(\varepsilon)$ denotes density of states(DOS), and at zero temperature, one has $\frac{1}{1+e^{\beta \varepsilon}}=\theta(-\varepsilon)$ and hence the above integral could be approximated as  (recall $E_a=\varepsilon_a-\bar f\sum_b n_b^0$)
\begin{eqnarray}
\label{comre}
&&\frac{\partial \mathcal N}{\partial \mu}=g(\mu-i\lambda_s+\bar f \sum_b n_b^0)(1-\bar f \frac{\partial \mathcal N}{\partial \mu}),\nonumber\\
&& (\mu=0, \text{one has } i\lambda_s=\bar f \sum_b n_b^0)\quad NK=\frac{g(0)}{1+\bar fg(0)}.
\end{eqnarray}

For specific heat, we first calculate energy of the system as ($\mu=0$)
\begin{eqnarray}
\bra E-\mu\mathcal N\ket =\bra E\ket&=&\frac{-\partial \ln \mathcal Z}{\partial \beta}\nonumber\\
&=&\frac{\lambda_s^2}{2\bar f}+\int g(\varepsilon_a) d\varepsilon_a \frac{E_a-\mu+i\lambda_s}{1+e^{\beta (E_a-\mu+i\lambda_s)}}\approx \frac{\lambda_s^2}{2\bar f}+\frac{\pi^2}{6} g(0) T^2.
\end{eqnarray}

At zero temperature, one has for the entropy 
\begin{eqnarray}
\mathtt S_{T=0}&=&0\quad\rightarrow \frac{\partial \mathtt S}{\partial \mathcal N}\Big|_{T=0}=0,\nonumber\\
\frac{\partial \mathcal N/\partial T}{\partial\mathcal N/\partial \mu}&=&\frac{\partial \mu}{\partial T}\Big|_{\mathcal N}= \frac{\partial \mathtt S}{\partial \mathcal N}\Big |_{T=0}=0\rightarrow \frac{\partial\mathcal N}{\partial T}\Big|_{T=0}=0,
\end{eqnarray}
where we have used a Maxwell relation in the second line.

Hence for specific heat
\begin{equation}
NC_{T=0}=\frac{\partial E}{\partial T}=\frac{\partial E(\mu,\beta,\lambda_s)}{\partial \lambda_s}\frac{\partial \lambda_s}{\partial T}\Big |_\mu+\frac{\pi^2}{3}g(0) T=\frac{\pi^2}{3}g(0) T.
\end{equation}

We see that while $C$ is unaffected by the interaction parameter $\bar f$, $K$ depends on $\bar f$ from \eqref{comre} and further  (introducing dimensionless $F=g(0)\bar f$)
\begin{equation}
\frac{\gamma}{K}=\frac{\pi^2}{3}(1+F)\sim \frac{U_0^2}{t_0^2}\gg 1
\end{equation}
bearing witness to the heavy Fermi liquid description.

\end{widetext}
 \bibliographystyle{apsrev4-1}
%

 \end{document}